\documentclass[prx,amsmath,amssymb, notitlepage, twocolumn,
nofootinbib,
superscriptaddress,
longbibliography,floatfix
]{revtex4-1}

\usepackage{amsmath}
\usepackage{tabularx,graphicx}
\usepackage{epstopdf}
\usepackage{graphicx}
\usepackage{latexsym}
\usepackage{amssymb}
\usepackage{amsmath}
\usepackage{color, colortbl}
\usepackage{psfrag}
\usepackage{bbm}
\usepackage{bm}
\usepackage{titlesec}
\usepackage{dsfont}
\usepackage{feynmp}
\usepackage{slashed}
\usepackage{multirow}
\newcommand{\beq}{\begin{eqnarray}}
	\newcommand{\eeq}{\end{eqnarray}}

\newcommand{\bsp}{\begin{split}}
	\newcommand{\esp}{\end{split}}

\usepackage{color}
\definecolor{darkblue}{rgb}{0.,0.,0.4}
\definecolor{darkred}{rgb}{0.5,0.,0.}
\definecolor{BlueViolet}{RGB}{138,43,226}
\definecolor{SkyBlue}{RGB}{30,144,255}
\definecolor{DarkGreen}{RGB}{0,100,0}
\usepackage[pdftex,colorlinks=true,linkcolor=darkblue,citecolor=blue,urlcolor=darkred]{hyperref}
\usepackage[normalem]{ulem}
\def\Z{\mathbb{Z}}

\usepackage{collectbox}

\usepackage{framed}

\begin{document}
	\title{Exploring critical systems under measurements and decoherence via Keldysh field theory}
	
	\author{Ruochen Ma}
	\affiliation{Perimeter Institute for Theoretical Physics, Waterloo, Ontario, Canada N2L 2Y5}
        \affiliation{Department of Physics and Astronomy, University of Waterloo, Waterloo, ON, N2L 3G1, Canada}

\begin{abstract}
We employ an $n$-replica Keldysh field theory to investigate the effects of measurements and decoherence on long distance behaviors of quantum critical states. We classify different measurements and decoherence based on their timescales and symmetry properties, and demonstrate that they can be described by $n$-replica Keldysh field theories with distinct physical and replica symmetries. Low energy effective theories for various scenarios are then derived using the symmetry and fundamental consistency conditions of the Keldysh formalism. We apply this framework to study the critical Ising model in both one and two spatial dimensions. In one dimension, we demonstrate that (1) measurements over a finite period of time along the transverse spin direction do not modify the asymptotic scaling of correlation functions and entanglement entropy, whereas (2) measurements along the longitudinal spin direction lead to an 
area law entangled phase. We also show that (3) decoherence noises over a finite time can be mapped to specific boundary conditions of a critical Ashkin-Teller model, and the entanglement characteristics of the resulting mixed state can be determined. For measurements and decoherence over an extensive time, we demonstrate that (4) the von Neumann entanglement entropy of a large subsystem can exhibit a (sub-)dominant logarithmic scaling in the stationary state for weak measurement (decoherence) performed in a basis that is symmetric under the Ising symmetry, but (5) reduces to an area law for measurements and decoherence in the longitudinal direction. Our results demonstrate that the Keldysh formalism is a useful tool for systematically studying the effects of measurements and decoherence on long-wavelength physics.

\end{abstract}

\maketitle

\tableofcontents

\section{Introduction}
\label{sec:intro}

The recent advancements in quantum devices \cite{2018nisq,2019Nature} have led to a renewed interest in the study of open quantum systems, where the dynamics is not solely governed by the Hamiltonian. Of particular interest is the class of monitored quantum systems, which experience both a deterministic unitary time evolution and stochastic state updates from measurements. The study of such systems has revealed the existence of a measurement-induced phase transition \cite{2019mipt1,2018mipt2,2019chan,2020gullans,2020jianyou,2020baochoi}, which causes a qualitative change in the entanglement properties of the system. More recently, the impact of decoherence has become a focal point of research due to the unavoidable interaction of realistic quantum devices with their surrounding environments, leading to noise-induced effects \cite{2023bao,2023Fan,2023decoxu,2023tim}.

From a physical perspective, the three aforementioned types of dynamics, namely Hamiltonian, measurement, and decoherence, have rather distinct effects on many-body quantum states: 
\begin{enumerate}
    \item A Hamiltonian, or a general unitary time evolution, typically results in the generation of quantum entanglement among microscopic degrees of freedom in the system. For instance, a local Hamiltonian usually produces entanglement in its ground-state within a correlation length. This correlation length can be as large as the entire system size near a quantum criticality.
    
    \item Local measurements with post-selection, on the other hand, project the state onto an eigenstate of a local observable, leading to a decrease in the entanglement between the measured microscopic component and the remaining system, as well as between the system and its environment. It is important to note that, despite the stochastic nature of the state update following measurement, an initial pure state of the system remains pure throughout the time evolution. At the end of the evolution, we obtain an ensemble of state trajectories that are each labeled by a particular sequence of measurement outcomes.
    
    \item Decoherence arises from the interaction between a quantum system and its surrounding environment, which generates quantum entanglement between the two. Because all observables are associated with the system and not the environment, the latter is effectively traced out. Therefore, generic decoherence processes increase the mixedness of a quantum system.
\end{enumerate}
Previous investigations have revealed that the interplay between these dynamics can lead to a diverse range of collective phenomena and phase transitions. However, the majority of these studies have been focused on particular lattice models and specific types of measurements or decoherence.

In this study, we aim to provide a description of the universal features of an open system characterized by a large correlation length, which can arise from any of the three aforementioned dynamics or their competition. For the sake of simplicity, we focus on the impact of local measurements or decoherence (“perturbations"), on the ground-state of a quantum Hamiltonian at criticality. Specifically, we examine two timescales of the perturbations: a finite time perturbation that is independent of the system size, and a perturbation over a period of time that is comparable to the system size. These two time scales are motivated by the following two questions respectively: (1) What is the nature of the quantum state resulting from measurements or decoherence on a critical ground-state? (2) What properties of the critical ground-state survive in the stationary state, when a critical Hamiltonian is in an environment with measurements or decoherence? In order to approach the long wavelength physics, in particular the ensemble-averaged correlation functions and entanglement entropy, we use a replicated Keldysh field theory \cite{2016Diehl,kamenev2023field} to describe the effect of measurements or decoherence.

The use of the Keldysh effective theory provides a valuable quantum field theory toolbox, which offers two key advantages: (1) It makes the microscopic and replica symmetries and fundamental consistency conditions of a density matrix manifest. (2) It enables the identification of the physically relevant degrees of freedom at low energies (IR). In general, our approach is to identify the internal symmetries of the system at microscopic scales (UV) and then construct the most general low energy effective theory that satisfies the symmetry, using IR degrees of freedom. The IR behaviors can then be deduced from this IR effective theory. 

As an example, we apply this approach to critical Ising model with a $\Z_2$ global symmetry in both one and two spatial dimensions. Specifically, we consider measurements and decoherence in either a $\Z_2$ even or odd basis. Our results show that, in one spatial dimension, after averaging over the entire ensemble of measurement outcomes: (1) measurements in a $\Z_2$ even basis over \emph{a finite period of time} do not alter the scaling behaviors of correlation functions and entanglement entropy compared to the initial critical state; (2) measurements in a $\Z_2$ odd basis cause the entanglement entropy to saturate to a constant for large subsystems. Furthermore, we find that different decoherence noises over a finite time can be mapped to distinct boundary conditions of a critical Ashkin-Teller model, and entanglement characteristics of the resulting mixed state, such as the $g$-function and the subsystem entropy, can be calculated accordingly. As an illustration, the mixed state arising from decoherence in the longitudinal direction is characterized by a value of $g=1/2$, and a subdominant logarithmic term in the second Renyi entropy, with a coefficient equal to $1/8$. On the other hand, when measured or decohered in a $\Z_2$ even basis, \emph{in the stationary state} the von Neumann entanglement entropy can still exhibit a logarithmic scaling for large subsystems\footnote{When referring to the entropy in the decoherence scenario, a volume law contribution is always subtracted throughout this study.}. Conversely, when measured or decohered in a $\Z_2$ odd basis, the stationary state exhibits an area law entanglement for arbitrarily small measurement/decoherence rate. These results are organized by their symmetry breaking patterns and are summarized in the boxes in Sec.~\ref{sec:details}. The discussion of the IR behaviours in two spatial dimensions follows a similar approach. Several physical setups that are analyzed in this study have been previously investigated in the literature \cite{2023ehud,2023linC,2022altman,2023jian}.

This paper is structured as follows. In Section \ref{sec:general}, we present the general framework of the replicated Keldysh effective theory. Specifically, in Section \ref{subsec:symmetry}, we introduce the Keldysh path integral for an open system undergoing a Markovian quantum dynamics, which describes decoherence. We also discuss two different microscopic symmetry conditions. Furthermore, in Section \ref{subsec:constraints}, we elaborate on the fundamental consistency conditions of the Keldysh formalism, which are due to consistency conditions of a density matrix. The effect of measurements is discussed in Section \ref{subsec:measurement}, where we re-write it as a Keldysh path integral using the quantum state diffusion framework \cite{2006QSDJ}. In Section \ref{subsec:nreplicas}, we provide a detailed discussion of the replica symmetries of various cases, in the entire space-time and on its boundary (the time slice where the measurements/decoherence are performed), serving as a guideline for our $n$-replica IR theory. We then apply this formalism to the critical Ising model in one and two spatial dimensions in Section \ref{sec:details}. Finally, in Section \ref{sec:discussion}, we present a summary of our work and discuss several open questions for future study.

\section{Generalities}
\label{sec:general}

In this section, I present an analysis of the effects of measurement and decoherence on an open quantum system through a replicated Keldysh field theory. For this purpose, it is assumed that the system is nearly critical, where the correlation length is much larger than the lattice spacing, to allow for a valid coarse-grained continuum description. Measurement and decoherence give rise to certain interactions in the effective field theory, and the universal long wave-length physics is studied in Section~\ref{sec:details} using standard techniques.

To define the question more precisely, let us denote the linear size of our system by $L$. In this work I will focus on two possible physical settings, distinguished by the time scales of the system being measured or experiencing decoherence:
\begin{enumerate}
    \item One can consider a critical quantum system initially in a pure state that undergoes measurement and decoherence for a finite time interval. It is essential to note that this interval is of the order of unity $[\sim O(1)]$ and does not depend on the size of the system. In the thermodynamic limit, interactions resulting from the perturbations are confined to a single time slice in the field theory description. Physical characteristics of the modified quantum state can be determined by studying correlation functions at this time slice.
    
    \item In the second scenario, measurement or decoherence persists over an extensive duration of time, typically of $O(L)$ in the thermodynamic limit. This setup enables an exploration of the properties of stationary states and response functions. Accordingly, interactions resulting from measurement or decoherence are included throughout the time evolution in the Keldysh field theory.  
\end{enumerate}

Examples for both scenarios will be provided in Sec.~\ref{sec:details}. The connection between the effect of measurement/decoherence and the boundary or defect properties has been pointed out in recent literature \cite{2022altman,2023xu}. As typical discussions about low energy physics, we constrain the form of the IR effective field theory based on (1) symmetry of the time evolution; (2) intrinsic consistency conditions of the Keldysh formalism, which emerge from the fundamental properties of a density matrix. To illustrate these constraints, we examine the Lindblad quantum master equation that describes decoherence in the next two subsections. For the case of measurement, a formalism for trajectory-averaged properties has been proposed in Ref.~\cite{2021buchhold,2022ladewig} for measurement-induced phase transitions of Dirac fermions, which will be reviewed in Sec.~\ref{subsec:measurement} and applied to more general cases in Sec.~\ref{subsec:nreplicas}.

\subsection{The Keldysh action and symmetries}

\label{subsec:symmetry}

The Keldysh functional integral formulates the time evolution of a density matrix. As an example, we start with a Lindblad quantum master equation that describes an open quantum system under decoherence,
\begin{equation}
    \frac{d}{dt} \rho = \mathcal{L} \rho =  -i[H,\rho] + \sum_\alpha \gamma_\alpha [2 L_\alpha \rho L_\alpha^\dagger - \{ L_\alpha^\dagger L_\alpha, \rho \}],
    \label{eq:mastereq}
\end{equation}
where the Liouvillian $\mathcal{L}$ acts on the density operator $\rho$ from both the ket and the bra sides. The \emph{quantum jump} operators $L_\alpha$ encode the dissipative couplings of the system with its environment. The (single replica) Keldysh partition function is defined as 
$Z_1 = \mathrm{tr}[\rho(t_f)]$, in which we take the trace of the density matrix at a time $t_f$. Upon the introduction of external source fields, the partition function plays the role of the generating functional for correlation functions. In terms of coherent state path integral, the non-trivial action of $\mathcal{L}$ on both sides of $\rho$ leads to a doubling of degrees of freedom, 
characteristic of the Keldysh formalism \cite{2009Kamenev,2016Diehl},
\begin{equation}
\begin{split}
    Z_1  = &\int \mathrm{D} \phi_+ \mathrm{D} \phi_- \mathrm{exp} (iS[\phi_+,\phi_-]), \\
    S[\phi_+,\phi_-]  = &\int_{t,x}^{t_f} \Bar{\phi}_+ i \partial_t \phi_+ - \Bar{\phi}_- i \partial_t \phi_- - \mathcal{L}[\phi_+,\phi_-],\\
    \mathcal{L}[\phi_+,\phi_-]  = & H_+ - H_- + i \sum_\alpha \gamma_\alpha [2 L_{\alpha,+} L_{\alpha,-}^\dagger \\
    & -  L_{\alpha,+}^\dagger L_{\alpha,+} -L_{\alpha,-}^\dagger L_{\alpha,-} ],
\end{split}
\label{eq:Keldyshaction}
\end{equation}
where $H_+ = H[\phi_+]$ etc., $\phi_+$ and $\phi_-$ represent the dynamical fields on the forward and backward branches of the Keldysh contour, respectively. We shall employ the notation $S_D$ to represent the decoherence action, which corresponds to the contribution from the quantum jump operator $L$ in the Keldysh action in Eq.~(\ref{eq:Keldyshaction}). 

Upon doubling the degrees of freedom in the Keldysh functional integral approach, it becomes evident that the Keldysh action may possess a doubled symmetry. Consider a Hamiltonian that is symmetric under a group $G$, as well as an initial pure state that shares this symmetry. In the absence of dissipative couplings or when all $L_\alpha$ are $G$ invariant, the Keldysh action would exhibit a symmetry of $G_+ \times G_-$, which act on $\phi_+$ and $\phi_-$, respectively. In terms of the density matrix, they separately operate on the ket and bra sides of $\rho$. When the system-environment interaction $L_\alpha$ transforms in a non-trivial representation under $G$, the presence of dissipative coupling reduces the symmetry of the Keldysh action to $G$. This corresponds to the diagonal subgroup of $G_+\times G_-$ that operates on the two sides of $\rho$ adjointly. 

In Sec.~\ref{sec:details} we explore both symmetry conditions of the Keldysh action. In Ref.~\cite{2012strongweak,2022turzillo}, the doubled symmetry is referred to as the \emph{strong} symmetry and the diagonal subgroup as the \emph{weak} symmetry. We adopt this nomenclature throughout our study.

\subsection{Consistency conditions}
\label{subsec:constraints}

The definition of the partition function leads to a series of consistency conditions that the Keldysh action must satisfy, which, together with the symmetries discussed in Sec.~\ref{subsec:symmetry} and their $n$-replica generalizations in Sec.~\ref{subsec:nreplicas}, will serve as the guiding principles for our low energy effective theory. To facilitate our analysis, we introduce a new set of variables,:
\begin{equation}
    \phi_c = \phi_+ + \phi_-,\quad \phi_q = \phi_+ - \phi_-.
\end{equation}
The two new fields are usually called the classical and quantum component in literature \cite{kamenev2023field}. Importantly, the Keldysh formalism imposes three consistency conditions.

(1) Conservation of probability: When $\phi_q=0$, the Keldysh action vanishes identically:
\begin{equation}
    S[\phi_c,\phi_q=0]=0.
    \label{eq:probconserve}
\end{equation}
Intuitively, in the case of $\phi_+ = \phi_-$ the action on the forward branch exactly cancels that on the backward part. More precisely, this requirement is due to the trace-preserving property of the Lindblad master equation Eq.~(\ref{eq:mastereq}) and ensures the normalization of the partition function, $Z_1 = \mathrm{tr}[\rho(t_f)] = 1$. 

(2) Hermiticity: The density operator $\rho$ should always be Hermitian during the time evolution. In this regard, we consider the path integral representation of matrix elements:
    \begin{equation}
    \begin{split}
        & \langle \phi_1 | \rho(t) | \phi_2 \rangle  =  \int^{\phi_+(t)=\phi_1}_{\phi_-(t)=\phi_2} \mathrm{D}\phi_\pm \, \mathrm{exp}(i S[\phi_+,\phi_-])\\
       =  &\langle \phi_2 | \rho(t) | \phi_1 \rangle ^*  =  \int^{\phi_+(t)=\phi_2}_{\phi_-(t)=\phi_1} \mathrm{D}\phi_\pm \, \mathrm{exp}(-i S^*[\phi_+,\phi_-]) \\
        = & \int_{\phi_-(t)=\phi_2}^{\phi_+(t)=\phi_1} \mathrm{D}\phi_\pm \, \mathrm{exp}(-i S^*[\phi_-,\phi_+]),
    \end{split}
    \end{equation}
    in which the last line is merely a change of integration variables. The two matrix elements calculated above should be equal at any time $t$, indicating that 
\begin{equation}
    S[\phi_c,\phi_q] = - S^*[\phi_c,-\phi_q].
    \label{eq:hermiticity}
\end{equation}
Given the involvement of complex conjugation, one may consider the Hermiticity condition as an effective time-reversal symmetry. Hereafter, we refer to this constraint as the $\Z_2^T$ symmetry. It is worth noting that in order for this constraint to be fulfilled, the presence of an anti-unitary symmetry at the microscopic level is not a requirement for the system.

(3) Non-negativity: The preservation of non-negativity of $\rho$ during the time evolution is a crucial consideration, particularly in the case of decoherence where the density operator becomes mixed. In the framework of Lindblad dynamics which describes the decoherence process, the non-negativity condition necessitates that all dissipation rates $\gamma_\alpha$ appearing in the master equation Eq.~(\ref{eq:mastereq}) are non-negative \cite{gorini1976completely}. This condition on the other hand also ensures the convergence of the Keldysh path integral. Additionally, this requirement is equivalent to the criterion that the decoherence must be completely positive when viewed as a quantum channel. For $\gamma_\alpha > 0$, one can see that decoherence pushes the density matrix towards its diagonal, which agrees with our physical expectations.

\subsection{Measurements}
\label{subsec:measurement}

In Ref.\cite{2021buchhold,2022ladewig}, a formalism for many body systems under continuous measurements was proposed, and used to study monitored fermion dynamics. In this subsection, we will summarize the results in a manner appropriate for our purposes. A brief derivation of the results can be found in Appendix.~\ref{app:qsdderivation}.

To achieve a continuum description, we investigate the scenario of weak measurement, where information on a local degree of freedom is acquired at a finite rate, causing continuous changes to the state \cite{2006weakmeasure,2002trajectory}. On the other hand, local measurements are extensively performed on the entire system. We stress that after a measurement, a pure state still remains pure. The system's collective behavior is captured through various correlation functions that are averaged over the ensemble of post-measurement states. This approach is akin to the calculation of observables in disordered systems, where one averages over the realizations of disorder. 

In the first scenario, where measurements are performed at one time slice $t = t_f$, the ensemble of final state is labeled by their outcomes at different measurement locations. The post-measurement density matrix of a specific outcome is represented as
\begin{equation}
    \begin{split}
    & \rho(t_f) =  V \rho_\Omega V^\dagger, \\
    V = & \mathrm{exp}[ -\Gamma \int_x  M(x)^2  +\sqrt{\Gamma}  \int_x W_{t_f}(x)  M(x) ],
    \label{eq:qsdfinite}
    \end{split}
\end{equation}
where $\Gamma>0$ is a (small) effective measurement strength, and $\rho_\Omega$ denotes the state before measurements. $M(x)$ is the normal-ordered measurement operator -- the observable $O(x)$ being measured, subtracted by its expectation value in the initial state $\rho_\Omega$. $W_{t_f}(x)$ is a Gaussian random variable that reflects the stochastic nature of the post-measurement state update. It has zero mean and $\overline{W_{t_f}(x)W_{t_f}(y)} = \delta(x-y) $, where the overline denotes averaging over the ensemble of measurement outcomes.

We also consider continuous measurements that extend over a long period of time.
Through time evolution, we generate an ensemble of pure state trajectories, each of which corresponds to a distinct sequence of measurement outcomes. To describe the stochastic evolution of the density operator, we adopt the quantum state diffusion framework \cite{gisin1992quantum,wiseman1993interpretation}. Conditioned on a particular trajectory, the state-update can be expressed as follows:
\begin{equation}
    \begin{split}
    \rho(t+d t) = & V_{d t} \rho(t) V_{d t}^\dagger, \\
    V_{d t} = & \mathrm{exp}[ -\int_x(i  H +\gamma  M_t(x)^2 ) d t \\ &+\sqrt{\gamma}  \int_x W_t(x)  M_t(x) ].
    \label{eq:qsdinfinite}
    \end{split}
\end{equation}
Here the local measurement operator is defined as $M_t(x) = O(x)- \mathrm{tr}[\rho(t) O(x)]$, where $O(x)$ is the Hermitian local operator being measured, subtracted by its expectation value before the measurement. The explicit dependence of $M_t(x)$ on the expectation value reflects the measurement feedback to the time evolution. The parameter $\gamma>0$ represents the measurement strength. $W_t(x)$ is again a Gaussian random variable with zero mean and $\overline{W_t(x)W_{t'}(y)} = d t \delta(x-y) \delta(t-t')$, where the overline denotes averaging over the ensemble of trajectories. The proposed time evolution in Eq.~(\ref{eq:qsdinfinite}) can be rationalized by taking the limit of $\gamma\rightarrow\infty$, corresponding to a projective measurement where the quantum state rapidly collapses onto an eigenstate of the measured operator $O$.

The time evolution with measurement can be expressed as a Keldysh path integral straightforwardly. For example, after the Keldysh doubling, the evolution operator $V$ in Eq.~(\ref{eq:qsdinfinite}) gives rise to an additional term in Eq.~(\ref{eq:Keldyshaction}), given by:
\begin{equation}
    S_M[\phi_\pm] = i [ \int_{x,t} d t\gamma  M_{t,+}^2 -  \sqrt{\gamma} W_t M_{t,+} + (+ \rightarrow -)].
    \label{eq:measurementinducedacttion}
\end{equation}
An observation is that averaging over the ensemble of state trajectories in a single replica formalism yields the Lindblad form described by Eq.(\ref{eq:Keldyshaction}) with $L_\alpha = O$. This observation can be interpreted as the averaging of the measurement outcome with the corresponding measurement probability, leading to the erasure of the outcome information and resulting in an equivalent decoherence process with a corresponding quantum jump operator.

\subsection{Replica field theory}
\label{subsec:nreplicas}
This subsection extends the previous discussions in the preceding sections to an $n$-replica theory. As previously stated in the Introduction, the Keldysh formalism of measurement and decoherence opens a powerful toolbox of modern quantum field theory. Specifically, symmetries at high energy scales must be preserved under RG flow to long distances. Thus, in a strongly interacting theory, one can examine the long-wavelength physics by writing a generic IR effective action that preserves the symmetry, disregarding the details of the RG flow. In any local quantum field theory, it is essential to differentiate between two notions of symmetry: the symmetry of the effective action in space-time (bulk), and the symmetry of the boundary condition, which, in this case, is the time slice $t = t_f$ \footnote{In this paper, we use the terms “boundary" and “$t=t_f$" interchangeably.}. In our forthcoming discussion of $n$-replica theory, this distinction becomes crucial.

(1) Consider first the measurement scenario, where we are interested in the $n$-replica partition function $Z_n = \mathrm{tr}[\overline{\rho(t_f)^{\otimes n}}]$, where the trace is taken over a tensor product of $n$ Hilbert spaces and the overline denotes ensemble average. As an example, when measurements are performed throughout the time evolution, the measurement action in Eq.~(\ref{eq:measurementinducedacttion}) becomes
\begin{equation}
\begin{split}
     S_M[\phi_\pm] & = i [ \int_{x,t} d t\gamma \sum_\alpha (M_{t,+}^{(\alpha)})^2\\
    &-  \sqrt{\gamma} W_t \sum_\alpha M_{t,+}^{(\alpha)} + (+ \rightarrow -)],
 \end{split}
 \label{eq:nthmeasurementinducedacttion}
\end{equation}
where $\alpha$ denotes the replica index. Crucially, the stochastic variable $W_t$ couples to all replicas and both branches for each replica in an identical fashion, analogous to the case in disordered systems where the disorder potential couples to all replicas identically. This reflects the fact that all replicas of a single trajectory undergo the same stochastic state update, and the label of this trajectory, i.e., the sequence of measurement outcomes, carries physical significance, as previously stated in the Introduction. The correspondence between the dynamics with measurements and the effects of static disorders was also noted in a previous study \cite{2020Jian}.

In the absence of measurement, the discussion in Sec.~\ref{subsec:symmetry} can be straightforwardly generalized to the $n$-replica case, revealing that the Keldysh-doubled effective action exhibits a global symmetry $(G_+^{\otimes n}\rtimes S_n)\times (G_-^{\otimes n}\rtimes S_n)$ \cite{2021Bao}. Here, the two $S_n$'s represent the permutation group of the $+$ and $-$ branches, respectively. Upon introducing the measurement action $S_M$ at UV energy scales, the most important information we can extract is about the UV internal symmetry of the theory. Of particular interest are two possibilities:
\begin{itemize}
    \item 
    We first consider the case where the measurement action $S_M$ preserves the strong $G_+ \times G_-$ symmetry discussed in Sec.~\ref{subsec:symmetry}. This implies that the stochastic variable $W$ transforms trivially under $G$, i.e. the measurement is in a local symmetric basis. The bulk of the theory preserves the $(G_+^{\otimes n}\rtimes S_n)\times (G_-^{\otimes n}\rtimes S_n)$ symmetry. However, the boundary condition at $t = t_f$ has only a reduced symmetry of $G^{\otimes n} \rtimes S_n$. This can be understood by noting that, upon taking the trace in the partition function (gluing the boundaries), we make the identification:
    \begin{equation}
        \begin{split}
            \phi_+^{(\alpha)}(t_f) = \phi_-^{(\alpha)}(t_f) = \phi_{(\alpha)}.
        \end{split}
        \label{eq:boundarycondition}
    \end{equation}
    Thus the boundary condition preserves only the simultaneous action on the identified boundaries, along with an $S_n$ permutation of replicas.

    \item In situations where measurements are conducted in a basis that transforms non-trivially under $G$, the stochastic variable $W$ is associated with a non-trivial $G$ representation. The measurements action breaks the strong $G_+\times G_-$ symmetry described in Sec.~\ref{subsec:symmetry} down to the weak subgroup. The bulk theory exhibits either the full $(G_+^{\otimes n}\rtimes S_n)\times (G_-^{\otimes n}\rtimes S_n)$ symmetry when the measurement is conducted over a short timescale ($O(1)$), or a reduced $G\times (S_n\times S_n)$ symmetry for an $O(L)$ time measurement. The reason behind this symmetry reduction is the fact that the stochastic variable $W$ couples identically to all the $2n$ Keldysh branches. On the other hand, for our specific interest, the measurement action $S_M$ is always present on time slice $t = t_f$. Upon making the identification Eq.~(\ref{eq:boundarycondition}) at time slice $t = t_f$, the boundary condition has a symmetry $G \times S_n$.
\end{itemize}

The examples discussed in Sec.~\ref{sec:details} will be organized based on their patterns of symmetry breaking.

(2) We now shift our attention to the scenario of decoherence. In this case, the $n$-replica partition function is defined as $Z_n = \mathrm{tr}[\rho(t_f)^n]$. \footnote{In the presence of both measurements and decoherence, a unified partition function can be defined as $Z_n = \mathrm{tr}[\overline{\rho(t_f)^n}]$. In this study, we consider the effects of measurements and decoherence separately for the sake of simplicity.} Its path integral representation is given by the Keldysh effective action in the Lindblad form in Eq.~(\ref{eq:Keldyshaction}), with an additional summation over the contributions of the $n$ replicas. The Lindblad action, which describes the system at high energy scales, features an \emph{intra-replica} coupling between the two branches. The underlying interpretation, as explained in the Introduction, is that decoherence erases the measurement outcomes (integrates out the environment) in the first place, and the ensemble representation of a mixed state does not have a preferred basis.

Similarly, we focus on two distinct symmetry conditions:
\begin{itemize}
\item First, we examine cases where the strong $G_+ \times G_-$ symmetry is preserved by the decoherence. Depending on the decoherence timescale, the bulk theory may possess either the full $(G_+^{\otimes n}\rtimes S_n)\times (G_-^{\otimes n}\rtimes S_n)$ symmetry for $O(1)$ time decoherence, or a reduced $(G_+\times G_-)^{\otimes n}\rtimes S_n$ symmetry for $O(L)$ time decoherence. Notably, the intra-replica branch coupling prohibits independent permutation of the two branches. In contrast, the decoherence action $S_D$ is always present at the final time slice, $t = t_f$. To compute the partition function in the case of decoherence, we need to specify the boundary condition
\begin{equation}
    \begin{split}
            &\phi_+^{(\alpha)}(t_f) = \phi_-^{(\alpha+1)}(t_f) = \phi_{(\alpha)}, \\
           &\phi_-^{(\alpha)}(t_f) = \phi_+^{(\alpha-1)}(t_f) = \phi_{(\alpha-1)}.
    \end{split}
    \label{eq:boundaryconditiond}
\end{equation}
The boundary condition exhibits a reduced $G^{\otimes n }\times\Z_n$ symmetry, where $\Z_n$ corresponds to the cyclic permutation of replicas.

\item In cases where the decoherence preserves only the weak $G$ symmetry, depending on the time scale of the decoherence, the bulk theory has either the full $(G_+^{\otimes n}\rtimes S_n)\times (G_-^{\otimes n}\rtimes S_n)$ symmetry, or a reduced $G^{\otimes n}\rtimes S_n$ symmetry for $O(L)$ time decoherence. At the boundary $t = t_f$, the symmetry is reduced to $G\times\Z_n$ by the boundary condition in Eq.~(\ref{eq:boundaryconditiond}).

\end{itemize}

All the symmetry conditions enumerated in this subsection are considered as internal symmetries of the system under measurement or decoherence at UV energy scales, which must be preserved along the RG flow. Besides the symmetries discussed above, an extra time reversal symmetry $\Z_2^T$ arising from the Hermiticity of the density matrix is imposed for both the measurement and the decoherence scenarios.

After presenting the general formalism and fundamental constraints in this section, we are now prepared to investigate the impact of measurement and decoherence on the long wave-length behavior of specific critical systems. Our strategy is to construct the most general IR effective theory that are allowed by the UV internal symmetries. The IR counterparts of the local measurement operator $M$ and the quantum jump operator $L$ can also be identified, guided by symmetry considerations. Moreover, the Keldysh action must satisfy the consistency conditions outlined in Section~\ref{subsec:constraints}.

\section{Examples: the Ising model}
\label{sec:details}

Our analysis focuses on a transverse field Ising model with a $\Z_2$ global symmetry, in one ($1d$) or two ($2d$) spatial dimensions, and is tuned to the vicinity of the Ising critical point. The $\Z_2$ charged operators include the spin $Z$ operator acting on each lattice site. The product of spin $X$ on each site serves as the symmetry generator.

\subsection{Finite time perturbations: one dimension}
\label{subsec:1dfinite}
To begin, we examine the scenario where the system is subjected to a perturbation, i.e. measurements or decoherence, for a duration of $O(1)$ time. We categorize the possible perturbations based on their symmetry properties. The Ising Hamiltonian exhibits a $\Z_2$ spin flip symmetry, which after the Keldysh doubling, gives rise to a $\Z_2^+ \times \Z_2^-$ symmetry in the long wavelength field theory. Specifically, the $\Z_2^+$ ($\Z_2^-$) symmetry acts on the bra (ket) side of the density matrix.

We first consider the effects of measurements. Using the Keldysh formalism, we prepare the density matrix before measurement by allowing a purely Hamiltonian dynamics to evolve for an infinitely long time, resulting in the projection of the system to the pure ground state. In this setting, the action induced by the measurement described in Eq.~(\ref{eq:qsdfinite}) is solely present at the final time slice $t_f$. The partition function can then be calculated by closing the time contour. In a single replica formalism, performing the trajectory-average leads to 
\begin{equation}
S_M = i \frac{\Gamma}{2} \int_{x} (M_{+}-M_{-})^2,
\label{eq:singlereplicameas}
\end{equation}
which vanishes identically when the time contour is closed and the ket side is identified with the bra side. This observation implies that weak measurements do not affect the scaling behaviors of correlation functions of local operators that are linear in the density matrix. Specifically, the correlation function $\overline{\langle O(0) O(x) \rangle} = \mathrm{tr}[\Bar{\rho}(t_f) O(0) O(x)]$ retains the same scaling form as it had in the pre-measurement state, where $O$ denotes a local operator. 

However, the situation becomes different when we consider quantities non-linear in density matrix, e.g. the famous von Neumann entropy. We primarily construct a theory with two copies of the density matrix, for the sake of simplicity. We begin by considering the following quantity (at time slice $t_f$, after measurements):
\begin{equation}
    \rho_2 = \overline{\rho \otimes \rho},
    \label{eq:measurementaction}
\end{equation}
which is a tensor product of two identical density matrices (conditioned on the same measurement outcome trajectories), and then averaged over the trajectory ensemble. This quantity can be constructed using a two-replica Keldysh effective theory, and the measurement-induced coupling can be expressed, at UV energy scales, as follows:
\begin{equation}
\begin{split}
    S_M= i\Gamma\int_x \{ \sum_{\sigma=\pm}\sum_{\alpha=1}^2 [M_{\sigma}^{(\alpha)}]^2  -\frac{1}{2} [\sum_{\sigma,\alpha} M_{\sigma}^{(\alpha)}]^2 \},
\end{split}
\label{eq:tworepmeasure}
\end{equation}
where $\alpha = 1,2$ is the replica index.

In order to compute the two-replica partition function $Z_2 = \mathrm{tr}[\overline{\rho(t_f)^{\otimes 2}}]$, it is necessary to glue the time slice $t_f$ with the boundary condition in Eq.~(\ref{eq:boundarycondition}).
It should be noted that the bulk far from the time slice $t_f$ is always described by an Ising conformal field theory (CFT). Depending on the basis in which the system is being measured, we consider two scenarios, distinguished by symmetry of the measurement operator.

(1) When measuring a system in a symmetric basis, such as along the $X$ basis, the measurement operator conserves the spin flip symmetry. In the case of two replicas, the internal symmetry of the UV theory described by Eq.~(\ref{eq:tworepmeasure}) reduces to $\Z_2^2 \times \Z_2^{(p)}$ at the boundary $t=t_f$. Here $\Z_2^{(p)}$ denotes the replica permutation symmetry. To capture the effects of measurement at low energies, we must enumerate all local interactions in the IR effective theory that are consistent with this symmetry. The dominant contribution to the IR effective theory can be expressed as
\begin{equation}
\begin{split}
    S_M^{IR} 
     = -i  \Gamma \epsilon_{(1)} \cdot \epsilon_{(2)}+ ...,
    \label{eq:measuresym}  
\end{split}
\end{equation}
where terms with higher scaling dimensions have been omitted. Meanwhile, the dominant contribution to the measurement operator $M$ at low energies can be identified as $M \sim \epsilon$. Several comments follow:
\begin{enumerate}
    \item The derivation of the IR effective action $S_M^{IR}$ from the UV theory in Eq.~(\ref{eq:tworepmeasure}) is generally not feasible due to the strongly-interacting nature of the underlying system. The only information we have is that $S_M^{IR}$ retains the global symmetry of the UV theory.
    
    \item An additional term proportional to $\epsilon_{(1)} + \epsilon_{(2)}$ in Eq.(\ref{eq:measuresym}) could also preserve the symmetry of the boundary condition. However, the $(1+1)d$ Ising model has an emergent symmetry -- the Kramers-Wannier duality which flips the sign of $M_{(1)}\sim\epsilon_{(1)}$ and $M_{(2)}\sim\epsilon_{(2)}$ simultaneously in Eq.~(\ref{eq:tworepmeasure}). As a consequence, this symmetry forbids the inclusion of the aforementioned coupling. It should be noted that this is a specific feature of the $(1+1)d$ Ising model\footnote{Strictly speaking, here the unperturbed UV fixed point is considered as an Ising CFT with an enlarged symmetry. In terms of a lattice model perspective, this corresponds to a scenario in which the measurement operator $M$ transforms nicely (with only a $-1$ factor) under the Kramers-Wannier duality even at high energy scales.}.

    \item It is worth noting that the interaction given in Eq.~(\ref{eq:measuresym}) preserves the $\Z_2^T$ time reversal symmetry, which acts on the $t = t_f$ slice as a complex conjugation. Note that according to our convention in Eq.(\ref{eq:hermiticity}), a minus sign obtained by $S_M^{IR}$ under $\Z_2^T$ implies time reversal invariance.  
\end{enumerate}
 
As the scaling dimension of $\epsilon$ in the $(1+1)d$ Ising CFT is $\Delta_\epsilon = 1$, $S_M^{IR}$ is irrelevant on the 1D time slice $t_f$. The full symmetry of the original UV theory will remain preserved in the IR, and the scaling of correlation functions and the Renyi entropy remains unchanged.

Before proceeding, let me briefly comment on the scenario where the operator $M$ severely breaks the Kramers-Wannier duality at UV energy scales. It is then necessary to include the perturbation $S_M^{IR} = -i \Gamma' [\epsilon_{(1)}+\epsilon_{(2)}]$ which is exactly marginal at the defect $t=t_f$. Consequently, the scaling of the Renyi entropy and correlation functions exhibits continuous variation depending on the measurement strength \cite{2010defectentro,2012renyidefect,2015defectentro}. However, if one performs an analytical continuation of the number of replicas to $n\to 1^+$, the coefficient $\Gamma'$ must vanish due to the reasoning around Eq.~(\ref{eq:singlereplicameas}) -- the single replica measurement action (for $O(1)$ time measurements) vanishes identically. Therefore, in the $n\to 1^+$ limit these scalings, such as the von Neumann entanglement entropy, would remain unaffected by this exactly marginal $\epsilon$ linear term, see Eq.~(\ref{eq:vnentropysym}). As a result,
\begin{framed}
Measurements performed in a symmetric basis over a finite period of time, in $(1+1)d$ Ising CFT do not affect the scaling properties of trajectory-averaged correlation functions linear or non-linear in the density operator, as well as that of the von Neumann entanglement entropy.  
\end{framed}

(2) In the case of measurements performed in a $\Z_2$ odd basis, such as the local $Z$ basis, for the two-replica case the symmetry at $t = t_f$ becomes $\Z_2 \times \Z_2^{(p)}$. At long distances the operator $M$ should be identified with the most relevant local operator with the same symmetry property, namely $M\sim\sigma$, the spin field. A similar enumeration yields the measurement-induced coupling for measurements performed in a $\Z_2$ odd basis
\begin{equation}
    S_M^{IR} = -i  \Gamma \sigma_{(1)}\cdot \sigma_{(2)},
    \label{eq:measureasym}
\end{equation}
where the most relevant contribution is retained.
Unlike the symmetric measurement, here we have a relevant perturbation on the time slice given $\Delta_\sigma = \frac{1}{8}$. In effect, we have two copies of Ising CFT coupled to each other on a line defect (the time slice $t= t_f$). Each copy acts as a symmetry-breaking field for the other -- therefore both copies would be cut open at the $1d$ defect $t=t_f$, and in total we have four decoupled halves. Given the measurement rate $\Gamma>0$ \footnote{In principle, the functional dependence of the IR coupling constant on the microscopic measurement rate $\Gamma$ can be complicated. Nonetheless, we anticipate that it will remain non-negative due to the physical expectation that measurements tend to project the density matrix onto its diagonal elements, as outlined in the Introduction. Moreover, this term has a clear UV correspondence in the measurement action at high energies if one substitutes $M\sim \sigma$ into Eq.~(\ref{eq:tworepmeasure}), where the coefficient is indeed the (positive) microscopic measurement rate. As a relevant coupling, we expect it to increase monotonically along the RG flow. Hence, we use the notation $\Gamma$ for both the microscopic measurement rate and the IR coupling constant.}, 
this coupling favors configurations in which spins at the edge of the four halves are parallel. 

How to characterize the trajectory ensemble under $\Z_2$ odd measurements? Cardy established \cite{cardy1986effect,cardy1989boundary} that there are only three universality
classes of boundary conditions for an Ising CFT on a
semi-infinite plane: fixed-up ($\uparrow$), fixed-down ($\downarrow$), and free
($f$). As discussed earlier, the two copies of Ising CFT are cut open at the line defect, and the spins on the edge of the four halves have a parallel orientation. Despite the naive expectation of spontaneous breaking of the $\Z_2$ spin flip subgroup of the boundary symmetry, we do not anticipate any genuine symmetry breaking in the $(1+0)d$ time slice at $t=t_f$. Therefore, each of the four halves has a fixed boundary condition, and the boundary state can be written symbolically as $|\uparrow\uparrow\uparrow\uparrow\rangle + |\downarrow\downarrow\downarrow\downarrow\rangle$.

\begin{figure}
\begin{center}
  \includegraphics[width=.35\textwidth]{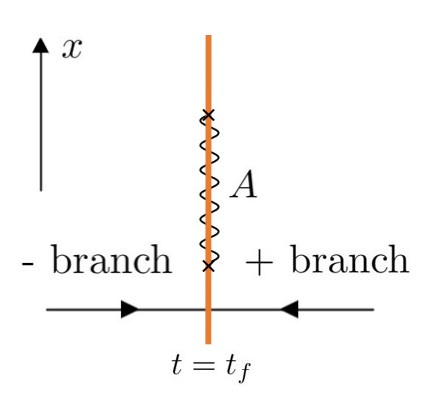} 
\end{center}
\caption{
Illustration of the boundary condition at $t=t_f$. Time
evolution of the density matrix is captured by two
Keldysh branches running horizontally, while the vertical line denotes the spatial dimension. The orange line marks the region where $S_M$ is non-vanishing, and the wavy line represents subsystem $A$ (the branch cut), where the entanglement entropy is calculated.
}
\label{Fig:measurement}
\end{figure}

This finding has significant implications on the entanglement entropy of the system subject to measurements. Consider a subsystem $A$ of length $l$, and let us compute the trajectory-averaged 2nd Renyi entropy defined as $S_A^{(2)} =-\overline{\mathrm{ln(tr}\rho_A^2)}$, where $\rho_A$ is the reduced density matrix of $A$. It has been demonstrated in Ref.~\cite{2009cardy,2005CardyCala} that the Renyi entropy can be computed on a space-time manifold with a boundary condition at the time slice $t = t_f$, such that (1) Inside $A$, the $n$ replicas are sewn together cyclically, leading to the identification in Eq.~(\ref{eq:boundaryconditiond}); (2) Outside $A$, each replica is sewn with itself, as that in Eq.~(\ref{eq:boundarycondition}). In the specific case of the Ising model, it can be checked that the boundary state inside and outside the subsystem $A$ remains the same, see Fig.~\ref{Fig:measurement}.\footnote{Upon examination of the boundary condition at $t = t_f$, the measurement action $S_M$ within subsystem $A$ is found to exhibit either a $G^{\otimes n}\times \Z_n$ symmetry (for measurements in a symmetric basis) or a $G\times \Z_n$ symmetry (for measurements in a $G$ non-trivial basis) at high energies. For $n=2$, the symmetries inside and outside of subsystem $A$ coincide.
}

In the case of measurements performed in a symmetric basis, the measurement induced coupling Eq.~(\ref{eq:measuresym}) on the defect is irrelevant (provided that the Kramers-Wannier duality is present). Thus the Renyi entropy remains the same as the case without the defect. Ref.~\cite{2009cardy} demonstrated that the Renyi entropy can then be related to scaling dimension of the twist field in Ising CFT (located at endpoints of the subsystem $A$), given by $S_A^{(2)} = \frac{1}{8} \mathrm{ln} l + O(1)$. In the limit $n \to 1^+$, the scaling of the ensemble-averaged von Neumann entanglement entropy $S_A=-\overline{\mathrm{tr}(\rho_A\mathrm{ln}\rho_A)}$, for any generic $\Z_2$ symmetric measurements, behaves as
\begin{equation}
    S_A = \frac{1}{6}\mathrm{ln}l+O(1).
    \label{eq:vnentropysym}
\end{equation}

In the scenario where measurements in a $\Z_2$ odd basis are carried out, the induced coupling by the measurement is relevant, thereby causing the system to be cut open at the $1d$ defect. The Renyi entropy is then calculated by the scaling dimension of certain boundary condition changing operator of the Ising boundary CFT. It is observed that, no matter inside or outside the subsystem $A$, the defect will always flow to the fixed boundary condition with the aforementioned boundary state. Consequently, the boundary condition changing operator at endpoints of $A$ is trivial. It is then deduced that the 2nd Renyi entropy saturates for large $l$, $S_A^{(2)}= O(1)$. In the limit of $n\to 1^+$, the validity of this area law remains for the von Neumann entanglement entropy.

\begin{framed}
 In a $(1+1)d$ Ising CFT, when measurements are performed in a $\Z_2$ odd basis over a finite time, the trajectory-averaged entanglement entropy of a subsystem saturates to a constant for sufficiently large subsystems.  \end{framed}

The effect of decoherence, which is described by a Keldysh effective action in the Lindblad form, can be studied similarly. The fundamental quantity of interest is the $n$-replica Keldysh partition function, $Z_n = \mathrm{tr}[\rho(t_f)^n]$. It is instructive to note a basic observation before examining specific decoherence. Let us denote the decoherence action by $S_D$, which is non-zero only at the time slice $t_f$. Consider the effect of decoherence over a finite time in a single replica formalism. When taking the trace in the partition function, $\phi_+(t_f)$ and $\phi_-(t_f)$ are identified. Conservation of probability, as given by Eq.~(\ref{eq:probconserve}), leads to the vanishing of the decoherence action $S_D$. This, combined with our previous statement, yields the following results:
\begin{framed}
    Weak measurement and decoherence over a finite period of time have no impact on the scaling behaviors of local operator correlation functions that are linear in the density operator.
\end{framed}
This is a general statement that extends to higher-dimensional systems as well. In the following discussion, we will concentrate on quantities that are non-linear in the density operator. Depending on the symmetry of the Lindbladian, we consider two possible scenarios.

(1) Let us first consider decoherence which preserves the strong $\Z_2^+ \times \Z_2^-$ symmetry, for example, a dephasing in the $X$ basis. Based on this symmetry, the low energy incarnation of the quantum jump operator should be identified as $L \sim \epsilon$. As previously mentioned, the boundary condition at $t = t_f$ preserves a $\Z_2^2\times \Z_2^{(p)}$ symmetry, as well as a $\Z_2^T$ time reversal invariance.

Now the aim is to enumerate all possible local interactions that maintain the $\Z_2^2\times \Z_2^{(p)}$ symmetry of the UV theory. The leading contribution is 
\begin{equation}
    S_D^{IR}  =  -i \gamma_D \int_x \epsilon_{(1)} \cdot \epsilon_{(2)}.
    \label{eq:xdecoherence}
\end{equation}
It is easy to see that this decoherence induced interaction is irrelevant on the $1d$ defect. As
a result, the scaling behavior of the decohered state is
identical to that without decoherence.\footnote{\label{footnote:commentonKW}
Again, by assuming the emergence of the Kramers-Wannier duality as an approximate symmetry, an exactly marginal contribution to the decoherence action given by $S_D^{IR} \sim -i\gamma_D' [\epsilon_{(1)} + \epsilon_{(2)}]$ is not included. However, if the quantum jump operator $L$ significantly breaks the Kramers-Wannier duality, this term should be taken into account. Incorporating this term induces a continuous variation in the scaling behavior of correlation functions and the Renyi entropy as the decoherence rate $\gamma_D'$ varies. However, it has no impact on these quantities in the $n\to 1^+$ limit where $\gamma_D'$ vanishes. As a result, the von Neumann entanglement entropy remains as $S_A = \kappa_1 l + \frac{1}{6}\mathrm{ln}l + O(1)$ where $l$ is the subsystem size.} For instance, the subsystem Renyi entropy scales as
\begin{equation}
    S_A^{(2)} = \kappa l +
    \frac{1}{8}\mathrm{ln} l + O(1), 
\end{equation}
where $\kappa$ is a scheme-dependent non-universal factor which can not be determined in the low energy field theory. It should be noted that we are currently dealing with a mixed state $\rho$ in the presence of decoherence, thus such a non-universal contribution from the configurational entropy is expected to appear.

(2) In the scenario of decoherence in a $\Z_2$ odd basis, such as a dephasing in the $Z$ direction, the Keldysh effective action preserves only a weak $\Z_2$ symmetry. At $t = t_f$, the UV symmetries of the boundary for two replicas are $\Z_2 \times \Z_2^{(p)}$ and the time reversal $\Z_2^T$. To construct an IR effective action, it is necessary to identify all local interactions that are consistent with this symmetry. The dominant contribution arises from 
\begin{equation}  
S_D^{IR} = -i \gamma_D \int_x  \sigma_{(1)} \cdot \sigma_{(2)},
    \label{eq:decoherence1danti}
\end{equation}
where the notation stems from the identification of fields at the boundary, $\sigma_+^{(1)}(t_f) = \sigma_-^{(2)}(t_f):=\sigma_{(1)}$ and $\sigma_-^{(1)}(t_f) = \sigma_+^{(2)}(t_f): = \sigma_{(2)}$. Meanwhile, at IR the quantum jump operator should be identified with the most relevant local operator that is $\Z_2$ odd, $L\sim\sigma$. As in the measurement scenario in Eq.~(\ref{eq:measureasym}), this interaction is relevant and flows to strong coupling at long distances. The non-negativity constraint in Eq.~(\ref{eq:Keldyshaction}) at high-energy scales\footnote{ The UV correspondence of the interaction in Eq.(\ref{eq:decoherence1danti}) becomes clear upon substitution of $L\sim \sigma$ into Eq.(\ref{eq:Keldyshaction}), with the coefficient $\gamma_D$ relating to the non-negative microscopic dissipation rate. }, as well as the physical expectation that decoherence pushes the density matrix towards its diagonal elements, suggest that the strength of decoherence $\gamma_D$ should be positive and increase monotonically as we flow towards lower energy scales. 

\begin{figure}
\begin{center}
  \includegraphics[width=.35\textwidth]{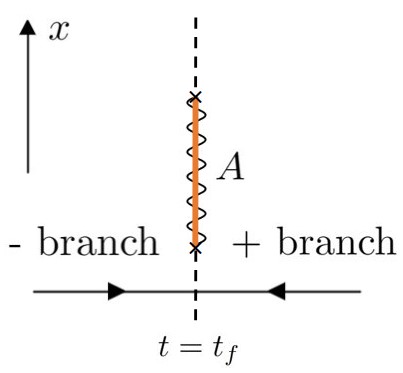} 
\end{center}
\caption{
Illustration of the boundary condition at $t=t_f$ for the case of decoherence. The orange line indicates the region where $S_D$ is present (inter-replica gluing), while the dashed line denotes a trivial defect (intra-replica gluing). It is noteworthy that the boundary conditions are different inside and outside subsystem $A$, where the entanglement entropy is evaluated.
}
\label{Fig:decoherence}
\end{figure}

Notably, despite the similarity between the expression for decoherence in Eq.~(\ref{eq:decoherence1danti}) and that for measurement, the Renyi entropy displays rather different features in the two scenarios. As illustrated in Fig.~\ref{Fig:decoherence}, while measurement induces the same boundary state inside and outside the subsystem $A$, the decoherence action $S_D$ leads to distinct boundary states in these regions. Here, $A$ refers to the subsystem for which the entropy is being computed.

Let us elucidate this point for the 2nd Renyi entropy. Inside the subsystem $A$, upon closing the time contour we have the boundary condition in Eq.~(\ref{eq:boundaryconditiond}), which yields the interaction term in Eq.~(\ref{eq:decoherence1danti}) at low energies, identical to the one derived in the measurement scenario in Eq.~(\ref{eq:measureasym}). Consequently, inside $A$, the four halves of the two copies of Ising CFT are cut open at the line defect, with spins on the edge being parallel to each other. In contrast, when closing the time contour outside the subsystem $A$, the resulting $S_D^{IR}$ vanishes. This is due to the fact that closing the time contour within each replica itself leads to $\sigma_+^{(\alpha)}(t_f)=\sigma_+^{(\alpha)}(t_f)$ at the line defect for both replicas, and the conservation of probability causes $S_D^{IR}$ to vanish identically. Therefore, outside $A$, the defect flows to a boundary state as if no defect were present.

What is the consequence of this observation on the subsystem entropy? Consider an Ising CFT on a complex plane with a line defect. Upon folding the system at the defect line, the defect of the Ising CFT maps to the boundary of a critical Ashkin-Teller model \cite{1991appliedcft} on a semi-infinite plane. The decoherence induced interaction inside the subsystem $A$ leads to a fixed boundary condition where the spins on either side of the defect are locked into a ferromagnetically aligned state, while outside $A$, the boundary state of the Ashkin-Teller model corresponds to a trivial defect in the Ising CFT. The corresponding boundary condition changing operator at endpoints of $A$ in the Ashkin-Teller boundary CFT has a dimension of $\Delta = 1/32$ \cite{1997affleck,1996oshikawa}. Therefore, the 2nd Renyi entropy of the decohered state is expressed as
\begin{equation}
    S_A^{(2)} = \kappa' l + \frac{1}{8}\mathrm{ln}l+O(1),
    \label{eq:entropyantibcc}
\end{equation}
where the $1/8$ factor corresponds to $2\times 2\Delta$ as we calculate the 2nd Renyi entropy using two copies of Ashkin-Teller boundary CFT. Upon taking the limit $n\to 1^+$, we find that the scaling of the von Neumann entanglement entropy follows an expression $S_A= \kappa_1'l + \frac{1}{16}\mathrm{ln}l+O(1)$.

The defect $g$-function, as defined in previous works \cite{cardy1986effect, cardy1989boundary, affleck1991universal} by $\ln g = (1-d/d\ln L)\ln Z_2(L)$ with $L$ being the linear system size, can be calculated accordingly. Physically, $g$ represents the overlap between the ground state of the Hamiltonian with periodic boundary condition and the boundary state. With symmet-
ric decoherence, the irrelevance of the defect coupling leads to two decoupled trivial line defects in the Ising CFT, resulting in $g = 1$.\footnote{ Even in cases where the Kramers-Wannier duality does not emerge and the contribution specified in Footnote.~\ref{footnote:commentonKW} is included, the exact marginality of this term ensures that $g$ retains its original value, i.e., $g=1$. }  Conversely, in the case of decoherence in an $\Z_2$ odd basis, as previously described, the boundary state $|\uparrow\uparrow\uparrow\uparrow\rangle+|\downarrow\downarrow\downarrow\downarrow\rangle$ has four copies of ferromagnetically aligned Ising fixed boundary conditions. The $g$-function is obtained by taking the product of $g$ for the four Ising layers, with an additional factor of 2 accounting for the multiplicity of $\uparrow$ and $\downarrow$. Thus, we have $g = 2 \times (1/\sqrt{2})^4 = 1/2$. Further breaking of the weak $\Z_2$ symmetry, such as decoherence in a direction slightly deviating from the $Z$ axis, would result in the removal of this degeneracy, resulting in $g=1/4$. Our results are consistent with those observed in a recent study \cite{2023tim}. Therefore, we have
\begin{framed}
    A $(1+1)d$ Ising CFT under decoherence over a finite period of time can be mapped to the boundary CFT of the critical Ashkin-Teller model, enabling us to analyze the entanglement characteristics in a precise manner. 
\end{framed}
This concludes our discussion on a finite time perturbation in $1d$.

\subsection{Finite time perturbations: two dimensions}

\label{subsec:finite2d}

Now we investigate the impact of an $O(1)$ time perturbation on a $(2+1)d$ system, specifically by considering the $(2+1)d$ Ising critical point as a case study. Throughout this subsection we focus on the two-replica formalism.

We begin with local measurements performed in a symmetric basis, where the associated local measurement operator is $Z_2$ even, and the action $S_M$ induced by the measurement preserves a $\Z_2^2\times \Z_2^{(p)}$ symmetry and the time reversal invariance due to Hermiticity. Note that in the case of $(2+1)d$ Ising model, the Kramers-Wannier duality is absent, resulting in the leading contribution to $S_M^{IR}$ being given by
\begin{equation}
    S_M^{IR} = -i \gamma_M (\epsilon_{(1)}+\epsilon_{(2)}).
    \label{eq:2+1symmetricmea}
\end{equation}
where now the coupling is present on a $(2+0)d$ defect in the $(2+1)d$ space-time. Utilizing $\Delta_\epsilon = 1.41$ in $(2+1)d$ Ising CFT \cite{2017bootstrap}, we establish that this coupling is relevant\footnote{Since this term is relevant as opposed to marginal, its impact cannot be disregarded even in the $n\to 1^+$ limit.}. Depending on the sign of the coupling constant $\gamma_M$, without further fine tuning, there can be two possible scenarios:
\begin{enumerate}
    \item In the scenario where $\gamma_M >0$, space-time of the system is cut open at the time slice $t = t_f$. the boundary at the cut is pushed to the ordinary boundary condition of a $(2+1)d$ critical Ising model, while the full symmetry of the UV theory is preserved in the IR. This scenario is expected to occur when the measurement is conducted, for example, in the local $X$ basis.

    \item For $\gamma_M < 0$, the coupling term again cuts the system at $t = t_f$, resulting in four open boundaries labeled as $(1,+),\,(1,-);\,(2,+),\,(2,-)$, corresponding to the replica indices and the branch labels, respectively. The first and last two boundaries flow separately towards the extraordinary boundary universality class, leading to the spontaneous breaking of the $\Z^2_2$ symmetry of the original UV boundary condition. A breaking of the $\Z_2^{(p)}$ replica permutation takes place when the spin orientations of the two pairs are not the same. This scenario is expected to occur when the measurement is performed in the local $Z_iZ_j$ basis.
\end{enumerate}

When the measurement is in a $\Z_2$ odd basis, $S_M$ at high energies preserves a $\Z_2\times \Z_2^{(p)}$ symmetry. Recall that the $\Z_2$ arises from the weak spin flip symmetry that acts adjointly on both sides of the density matrix. At low energies, the dominant contribution to $S_M^{IR}$ is again given by Eq.~(\ref{eq:measureasym}), with the local measurement operator identified with the Ising spin $\sigma$ at long distances. Since $\Delta_\sigma = 0.52$ in $(2+1)d$, $S_D^{IR}$ is relevant on the plane defect. As each replica can be viewed as a symmetry breaking field for the other, this coupling is expected to cut each replica into two halves. The four open boundaries at the plane defect are at the normal fixed point \cite{burkhardt1987surface,bray1977critical,cardy1996scaling}, with ferromagnetic alignment of spins on these boundaries, as $\Gamma>0$. This causes the $\Z_2$ spin flip symmetry to be spontaneously broken at the plane defect. The trajectory-averaged correlation functions can be calculated accordingly. For example, we expect the connected correlation function
\begin{equation}
    \overline{\langle Z(0) Z(x) \rangle} - \overline{\langle Z(0) \rangle \langle Z(x) \rangle }\sim O(1)\,\,\mathrm{const},
\end{equation}
where $Z$ is the spin $Z$ operator (or a generic $\Z_2$ odd operator). This long range order reveals the spontaneous $\Z_2$ symmetry breaking in the trajectory ensemble after measurement.

In the presence of decoherence, the effect on critical systems can be analyzed in a similar manner.
In the case of decoherence occurring in a symmetric basis, the resultant $S_D^{IR}$ is characterized by a $\Z_2^2\times \Z_2^{(p)}$ symmetry of the UV boundary condition. The leading contribution to $S_D^{IR}$ is given by $S_D^{IR} = -i \gamma_D (\epsilon_{(1)}+\epsilon_{(2)})$, which is relevant on a $(2+0)d$ defect in $(2+1)d$ spacetime. Long-wavelength behaviours are analyzed in the same manner as those below Eq.~(\ref{eq:2+1symmetricmea}).

In the case of decoherence occurring in a local $\Z_2$ odd basis, the boundary condition at $t=t_f$ has a $\Z_2 \times \Z_2^{(p)}$ symmetry at UV. In a two-replica scheme, the leading contribution of the decoherence action is again given by Eq.~(\ref{eq:decoherence1danti}), which drives the defect plane $t=t_f$ to a normal fixed point for both replicas. The $\Z_2$ spin flip symmetry is broken spontaneously, which is manifested by a non-zero long-range order: 
\begin{equation}
    \mathrm{tr}[\rho(t_f) Z(0) \rho(t_f) Z(x)]/\mathrm{tr}[\rho(t_f)^2] \sim \mathrm{const}.
\end{equation}
Additionally, given the expectation that decoherence tends to suppress the off-diagonal elements of the density matrix, and the correspondence between the coupling $\gamma_D$ and the microscopic dissipation rate at high energies, it is reasonable to assume that $\gamma_D$ remains positive in the IR, and consequently the $\Z_2^{(p)}$ symmetry is preserved. Intriguingly, akin to the $1d$ case, when calculating subsystem Renyi entropy, distinct time contour closings at the plane defect lead to distinct boundary states inside and outside the subsystem $A$. Consequently, we have a boundary condition changing line defect at the edge of $A$. A comprehensive study of this line defect remains an interesting unresolved issue for future investigations.

\subsection{The stationary state: one dimension}
\label{subsec:stationary1d}

In this subsection, we investigate an alternative physical scenario, specifically a critical Hamiltonian subjected to a perturbation (measurement or decoherence) for a duration of $O(L)$ time comparable to the system size, resulting in the system achieving a stationary state. To exemplify this scenario, we utilize the critical Ising model in $(1+1)d$ and examine the entanglement characteristics of the stationary state.

We begin with a continuous weak measurement in a local $\Z_2$ symmetric basis, for example, in the $X$ direction. The $n$-replica measurement action at the UV energy scales in Eq.~(\ref{eq:nthmeasurementinducedacttion}) preserves a $(\Z_2^{\otimes n}\rtimes S_n)\times (\Z_2^{\otimes n}\rtimes S_n)$ symmetry in the bulk space-time. The most general IR effective action allowed by the UV internal symmetry can be expressed as
\begin{equation}
\begin{split}
    S_M^{IR} = & -i\int_{x,t}  \sum_{\alpha \ne \beta}[ \gamma_1 \epsilon_+^{(\alpha)}\cdot \epsilon_+^{(\beta)} +\gamma_1^* \epsilon_-^{(\alpha)}\cdot \epsilon_-^{(\beta)} ] \\
    & -i\int_{x,t} \gamma_2 \sum_{\alpha,\beta} \epsilon_+^{(\alpha)}\cdot \epsilon_-^{(\beta)},  
\end{split}
\label{eq:conformalpert}
\end{equation}
where $\gamma_2$ is real due to the $\Z_2^T$ symmetry\footnote{
In this subsection we ignore the term linear in $\epsilon$ by assuming the preservation of the Kramers-Wannier duality. If this assumption is relaxed, a term $S_M^{IR} = -i \int_{x,t} \sum_\alpha [\gamma_1\epsilon_+^{(\alpha)} + \gamma_1^*\epsilon_-^{(\alpha)}]$ needs to be taken into account, which is relevant and results in an area law entangled phase.
}. Notably, this is a marginal interaction in $(1+1)d$ Ising CFT. Here, we determine the long-distance behavior through a perturbative RG.

Using Jordan-Wigner transformation \cite{sachdev1999quantum}, we map the Ising model to a free Majorana fermion and obtain the full low energy effective theory,
\begin{equation}
    \begin{split}
        S^{IR} = & \int_{x,t} \sum_\alpha \Bar{\chi}_{\alpha} i\slashed{\partial}\chi_{\alpha} - \Bar{\psi}_{\alpha} i\slashed{\partial}\psi_{\alpha} + S_M^{IR},\\
         S_M^{IR} = & -i\int_{x,t}  \sum_{\alpha \ne \beta}( \gamma_1 \Bar{\chi}_\alpha\chi_\alpha \Bar{\chi}_\beta\chi_\beta +\gamma_1^* \Bar{\psi}_\alpha\psi_\alpha \Bar{\psi}_\beta\psi_\beta ) \\
    & -i\int_{x,t} \gamma_2 \sum_{\alpha,\beta} \Bar{\chi}_\alpha\chi_\alpha \Bar{\psi}_\beta\psi_\beta, 
    \end{split}
    \label{eq:isinglongmeasure}
\end{equation}
in which $\chi$ and $\psi$ correspond to Majorana fermions on the forward and backward branches, respectively. The energy field $\epsilon$ is mapped to the Majorana mass \cite{2019Chong}. The beta function at the one-loop level is given by\footnote{The propagator of $\chi$ on the forward branch has an $i\eta$ prescription opposite to that of $\psi$ in Keldysh field theory. This is crucial when performing the Wick rotation.}
\begin{equation}
    \begin{split}
& \frac{d\gamma_1}{d\mathrm{ln}\mu} = -\frac{4(n-2)i}{2\pi}\gamma_1^2+\frac{ n i}{2\pi}\gamma_2^2, \\
& \frac{d\gamma_2}{d\mathrm{ln}\mu} = \frac{4(n-1)i}{2\pi}\gamma_1^*\gamma_2 - \frac{4(n-1)i}{2\pi}\gamma_1\gamma_2,
\end{split}
\label{eq:initialRG}
\end{equation}
where $\mu$ denotes a cutoff scale.

\begin{figure}
\begin{center}
  \includegraphics[width=.45\textwidth]{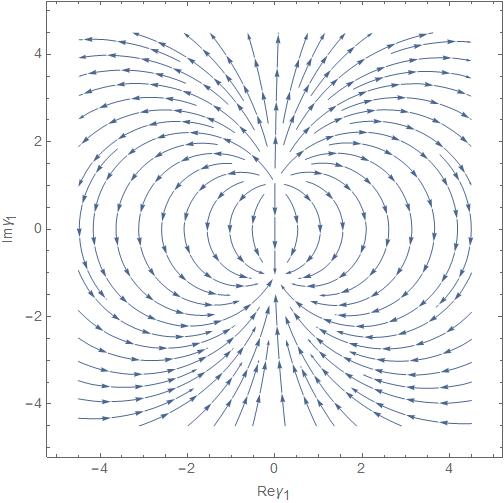} 
\end{center}
\caption{
RG flow of the parameter $\gamma_1$ in the complex plane. In the limit of $n\to 1^+$, the flow of $\gamma_2$ is much slower than that of $\gamma_1$. Hence, when discussing the flow of $\gamma_1$, we can treat $\gamma_2$ as a constant. The plot shows the RG flow for $n=1.05$ and $\gamma_2=2$.
}
\label{Fig:flow}
\end{figure}

Our primary focus is on the $n \to 1$ regime, where we can gain insight into the scaling behavior of the von Neumann entanglement entropy. We assume that $\gamma_1$ and $\gamma_2$ are real and positive at high energies, as they can be identified with the microscopic measurement rate at the UV scale. In the $n\to 1$ limit, $\gamma_1$ rapidly flows to its fixed point value $\gamma_1^{\mathrm{fp}}\sim -i\gamma_2/2$, as depicted in Fig.~\ref{Fig:flow}. On the other hand, the value of $\gamma_2$ drifts slowly according to
\begin{equation}
    \frac{d \gamma_2}{d\mathrm{ln}\mu}|_{\gamma_1 = \gamma_1^{\mathrm{fp}}}=-\frac{4(n-1)}{2\pi}\gamma_2^2.
    \label{eq:slowdrift}
\end{equation}
Therefore, $\gamma_2$ is a marginally relevant perturbation. Physically, we anticipate that the system flows to an Ising disordered/ordered phase (depending on the basis of the measurements) at the longest wavelength, with a vanishing drifting velocity in the replica limit $n\to 1^+$.\footnote{We can physically understand this by noting that when $n=1$, the $\gamma_1$ term disappears, and the $\gamma_2$ term reduces to an exactly marginal deformation $\sim\epsilon_1\epsilon_2$ of the Ashkin-Teller model.}

This observation suggests intriguing scaling behaviors of the trajectory-averaged entanglement entropy. Specifically, the RG flow in Eq.(\ref{eq:slowdrift}) implies that, for $n$ sufficiently close to $1$, a correlation length much larger than any subsystem size $l$ can always be generated. Therefore, if a critical Ising Hamiltonian is placed in an environment with $\Z_2$ symmetric measurements at a small measurement rate, after an extensive period of time the von Neumann entanglement entropy of a subsystem should still exhibit a logarithmic scaling with respect to $l$ for very large subsystems. Alternatively, if a product state with zero entanglement is initially prepared and the system is subsequently evolved according to the combined dynamics described in Eq.~(\ref{eq:isinglongmeasure}), the entanglement entropy of a subsystem is expected to show a logarithmic growth over time, eventually reaching a size-dependent value for sufficiently large time. The effective central charge, which governs the logarithmic growth of entanglement, may vary continuously with the measurement rate $\gamma_2$ at microscopic scales. Indeed, this phenomenon was recently observed in a numerical study \cite{2021Turkeshi}.

In the case of measurements conducted in a $\Z_2$ odd basis, for instance, the local $Z$ basis, at high energies the $n$-replica measurement action in Eq.~(\ref{eq:nthmeasurementinducedacttion}) preserves a $\Z_2\times S_n \times S_n$ symmetry and a $\Z_2^T$ time reversal. The leading contribution at the low energies is 
\begin{equation}
    \begin{split}
        S_M^{IR} = & -i\int_{x,t} \sum_{\alpha \ne \beta}\{ [\gamma_1\sigma_+^{(\alpha)}\cdot\sigma_+^{(\beta)}+\gamma_1^*\sigma_-^{(\alpha)}\cdot\sigma_-^{(\beta)}] \\
        & + \sum_{\alpha, \beta} \gamma_2 \sigma_+^{(\alpha)}\cdot\sigma_-^{(\beta)} \}.
    \end{split}
\end{equation}
This is a relevant perturbation that would drive the state to an area law entangled phase where the $\Z_2$ spin flip symmetry is spontaneously broken in a ferromagnetic manner\footnote{In light of the tendency for measurement to project the density matrix onto its diagonal and the correspondence at high energies between the coupling $\gamma_2$ and the microscopic measurement rate, it is reasonable to hypothesize that $\gamma_2$ retains its positivity along the RG flow. This, in turn, implies that the magnetizations of different replicas are ferromagnetically aligned, preserving the $S_n\times S_n$ replica permutation symmetry at low energies.}.
\begin{framed}
    For a critical Ising chain subject to weak measurements in a local $\Z_2$ symmetric basis, it is anticipated that the logarithmic scaling of the von Neumann entanglement entropy can still be observed for large subsystems in the stationary state (provided that the Kramers-Wannier duality is present). On the other hand, when measurements are conducted in a $\Z_2$ odd basis, the stationary state is in a phase characterized by spontaneous breaking of the $\Z_2$ spin flip symmetry and an area law entanglement entropy.
\end{framed}

Consideration can also be given to the characteristics of a quantum system with a critical Hamiltonian that undergoes decoherence over an extensive period of time. When the decoherence preserves the strong $\Z_2^+ \times \Z_2^+$ symmetry, the $n$-replica Lindblad action exhibits a $(\Z_2 \times \Z_2)^{\otimes n}\rtimes S_n$ symmetry and is time reversal invariant. The IR effective action may include all local interactions that are allowed by this symmetry. Among these couplings, the dominant contribution is given by
\begin{equation}
\begin{split}
    S_D^{IR}& =  -i\int_{x,t}  \sum_{\alpha \ne \beta}[ \gamma_1\epsilon_+^{(\alpha)}\cdot \epsilon_+^{(\beta)} + \gamma_1^*\epsilon_-^{(\alpha)}\cdot \epsilon_-^{(\beta)} ] \\
    & -i\int_{x,t} [\gamma_2 \sum_{\alpha\ne\beta} \epsilon_+^{(\alpha)}\cdot \epsilon_-^{(\beta)} + \gamma_3\sum_\alpha \epsilon_+^{(\alpha)}\cdot \epsilon_-^{(\alpha)}],  
\end{split}
\end{equation}
where $\gamma_2$ and $\gamma_3$ are real and positive parameters at high energies. By applying a Jordan-Wigner duality, it can be demonstrated that the one-loop beta function for the coupling constants is as follows:
\begin{equation}
    \begin{split}
 \frac{d\gamma_1}{d\mathrm{ln}\mu} = & -\frac{4(n-2)i}{2\pi}\gamma_1^2+\frac{ (n-2) i}{2\pi}\gamma_2^2+\frac{ 2 i}{2\pi}\gamma_2\gamma_3, \\
 \frac{d\gamma_2}{d\mathrm{ln}\mu} = & \frac{4(n-2)i}{2\pi}\gamma_1^*\gamma_2 - \frac{4(n-2)i}{2\pi}\gamma_1\gamma_2 \\
 & + \frac{4 i}{2\pi}\gamma_1^*\gamma_3 - \frac{4 i}{2\pi}\gamma_1\gamma_3, \\
\frac{d\gamma_3}{d\mathrm{ln}\mu} = & \frac{4(n-1)i}{2\pi}\gamma_1^*\gamma_2 - \frac{4(n-1)i}{2\pi}\gamma_1\gamma_2.
\end{split}
\end{equation}
In the replica limit, the values of $\gamma_1$ and $\gamma_2$ quickly approach their fixed point values, with $\gamma_1^{\mathrm{fp}}\sim -i\gamma_3/2$ and $\gamma_2^{\mathrm{fp}}\sim \gamma_3$, while the slow parameter $\gamma_3$ flows according to
\begin{equation}
    \frac{d \gamma_3}{d\mathrm{ln}\mu}|_{\gamma_1 = \gamma_1^{\mathrm{fp}}}^{\gamma_2 = \gamma_2^{\mathrm{fp}}}=-\frac{4(n-1)}{2\pi}\gamma_3^2.
\end{equation}
The analysis of the scaling behaviour of the von Neumann entanglement entropy is therefore the same as in the measurement scenario discussed below Eq.~(\ref{eq:slowdrift}).

When considering decoherence that only preserves the weak $\Z_2$ symmetry, such as dephasing in the $Z$ direction, the $n$-replica Lindblad action in Eq.~(\ref{eq:Keldyshaction}) exhibits a $\Z_2^{\otimes n}\rtimes S_n$ symmetry at high energies. The leading contribution to the quantum jump operator $L$ at low energies is $L\propto \sigma$. Based on this symmetry, the dominant contribution to the IR effective theory can be expressed as
\begin{equation}
    S_D = -i\gamma_D\int_{x,t} \sum_\alpha\sigma_+^{(\alpha)}\cdot\sigma_-^{(\alpha)}.
\end{equation}
At the UV scale, the coupling constant $\gamma_D$ can be identified as the dissipation rate, and thus it should be positive due to non-negativity. The decoherence action is relevant and leads to the spontaneous breaking of the $\Z_2^{\otimes n}$ symmetry, resulting in an area law entanglement after subtracting the volume law piece. Additionally, the $S_n$ replica permutation symmetry may also be broken if the magnetizations in different replicas have opposite signs. 

\begin{framed}
    For a critical Ising chain subject to local decoherence in a $\Z_2$ symmetric basis, it is anticipated that the subleading logarithmic scaling of the von Neumann entanglement entropy will persist for large subsystems in the stationary state, as long as the Kramers-Wannier duality is present. On the other hand, with decoherence in a $\Z_2$ odd basis, the stationary state is in a phase characterized by spontaneous breaking of the spin flip and replica permutation symmetries, as well as an area law entanglement (after subtracting the leading volume law piece arising from the mixedness of the state).
\end{framed}

\section{Discussions}
\label{sec:discussion}

In this study, we utilized a replicated Keldysh effective field theory to investigate the effects of measurements and decoherence on critical systems. Specifically, we examined both finite-time measurements/decoherence and the possible stationary state properties. Our results suggest that scalings of correlation functions of local operators that are linear in the density matrix remain unaffected by measurement and decoherence over a finite period of time. To analyze higher moments in the density matrix, we carefully distinguished the symmetry of an $n$-replica theory in various situations, both in the bulk and on the boundary. The low energy effective theory can be derived based on this symmetry, and the fundamental consistency conditions of the Keldysh path integral. We then applied this framework to the critical Ising model in one and two spatial dimensions, and low-energy behaviors under different perturbations, such as IR symmetry breaking patterns and correlation functions, were discussed. In one spatial dimension, we also explicitly calculated certain entanglement characteristics and discussed their connection to recent numerical studies.

We end with some open directions:
\begin{enumerate}
    \item
    One relevant issue is to define and classify topological phases in open quantum systems, such as the Symmetry-Protected Topological (SPT) phases \cite{2013SPTchen}, using a Keldysh effective field theory, and to investigate the implications of this non-trivial topology on various observables. A concrete example of driven-dissipative Chern insulator has been discussed in a remarkable study \cite{2020drivenchern}. A natural conjecture is, when a unique pure stationary state with a non-zero dissipative gap is present in the open quantum dynamics, one can define an SPT phase for this stationary state. Furthermore, it is expected that the classification of these phases will match that of the recently proposed Average Symmetry-Protected Topological phases \cite{2022aspt}. We will develop the theory of such phases in more detail in a forthcoming work.

    \item 
    Another open question is how to define and characterize topological orders for ensembles and mixed states \cite{2023bao,2023Fan}, or in non-unitary dynamics. This raises two fundamental questions: (1) how to define and determine the stability of conventional ground-state topological orders under local measurements and decoherence, and (2) can there exist topologically ordered states in the stationary state of non-equilibrium dynamics? Is it possible for there to be open system/non-unitary topological orders that lack a ground-state counterpart? These questions pertain to the two distinct timescales addressed in this study, and a comprehensive investigation of this matter is reserved for future work.
    
\end{enumerate}

\begin{acknowledgements}
This manuscript documents the author's attempt to comprehend a rapidly developing field, during which his mentors included Yin-Chen He, 
Leonardo A. Lessa, Tsung-Cheng Lu, Han Ma, Shengqi Sang,
Chong Wang, Jinmin Yi, Xuzhe Ying and Liujun Zou. The author is grateful to Zhen Bi, Meng Cheng, Timothy H. Hsieh, Tsung-Cheng Lu, Alex Turzillo, Chong Wang and Liujun Zou for their feedback on an earlier version of the manuscript. The author acknowledges supports from the Natural Sciences and Engineering Research Council of Canada(NSERC) through Discovery Grants. Research at Perimeter Institute is supported in part by the Government of Canada through the Department of Innovation, Science and Industry Canada and by the Province of Ontario through the Ministry of Colleges and Universities.

\end{acknowledgements}

\bibliography{ref.bib}

\begin{thebibliography}{53}%
\makeatletter
\providecommand \@ifxundefined [1]{%
 \@ifx{#1\undefined}
}%
\providecommand \@ifnum [1]{%
 \ifnum #1\expandafter \@firstoftwo
 \else \expandafter \@secondoftwo
 \fi
}%
\providecommand \@ifx [1]{%
 \ifx #1\expandafter \@firstoftwo
 \else \expandafter \@secondoftwo
 \fi
}%
\providecommand \natexlab [1]{#1}%
\providecommand \enquote  [1]{``#1''}%
\providecommand \bibnamefont  [1]{#1}%
\providecommand \bibfnamefont [1]{#1}%
\providecommand \citenamefont [1]{#1}%
\providecommand \href@noop [0]{\@secondoftwo}%
\providecommand \href [0]{\begingroup \@sanitize@url \@href}%
\providecommand \@href[1]{\@@startlink{#1}\@@href}%
\providecommand \@@href[1]{\endgroup#1\@@endlink}%
\providecommand \@sanitize@url [0]{\catcode `\\12\catcode `\$12\catcode
  `\&12\catcode `\#12\catcode `\^12\catcode `\_12\catcode `\%12\relax}%
\providecommand \@@startlink[1]{}%
\providecommand \@@endlink[0]{}%
\providecommand \url  [0]{\begingroup\@sanitize@url \@url }%
\providecommand \@url [1]{\endgroup\@href {#1}{\urlprefix }}%
\providecommand \urlprefix  [0]{URL }%
\providecommand \Eprint [0]{\href }%
\providecommand \doibase [0]{http://dx.doi.org/}%
\providecommand \selectlanguage [0]{\@gobble}%
\providecommand \bibinfo  [0]{\@secondoftwo}%
\providecommand \bibfield  [0]{\@secondoftwo}%
\providecommand \translation [1]{[#1]}%
\providecommand \BibitemOpen [0]{}%
\providecommand \bibitemStop [0]{}%
\providecommand \bibitemNoStop [0]{.\EOS\space}%
\providecommand \EOS [0]{\spacefactor3000\relax}%
\providecommand \BibitemShut  [1]{\csname bibitem#1\endcsname}%
\let\auto@bib@innerbib\@empty
\bibitem [{\citenamefont {{Preskill}}(2018)}]{2018nisq}%
  \BibitemOpen
  \bibfield  {author} {\bibinfo {author} {\bibfnamefont {John}\ \bibnamefont
  {{Preskill}}},\ }\bibfield  {title} {\enquote {\bibinfo {title} {{Quantum
  Computing in the NISQ era and beyond}},}\ }\href {\doibase
  10.22331/q-2018-08-06-79} {\bibfield  {journal} {\bibinfo  {journal}
  {Quantum}\ }\textbf {\bibinfo {volume} {2}},\ \bibinfo {pages} {79} (\bibinfo
  {year} {2018})},\ \Eprint {http://arxiv.org/abs/1801.00862} {arXiv:1801.00862
  [quant-ph]} \BibitemShut {NoStop}%
\bibitem [{\citenamefont {{Arute}}\ \emph {et~al.}(2019)\citenamefont
  {{Arute}}, \citenamefont {{Arya}}, \citenamefont {{Babbush}}, \citenamefont
  {{Bacon}}, \citenamefont {{Bardin}}, \citenamefont {{Barends}}, \citenamefont
  {{Biswas}}, \citenamefont {{Boixo}}, \citenamefont {{Brandao}}, \citenamefont
  {{Buell}}, \citenamefont {{Burkett}}, \citenamefont {{Chen}}, \citenamefont
  {{Chen}}, \citenamefont {{Chiaro}}, \citenamefont {{Collins}}, \citenamefont
  {{Courtney}}, \citenamefont {{Dunsworth}}, \citenamefont {{Farhi}},
  \citenamefont {{Foxen}}, \citenamefont {{Fowler}}, \citenamefont {{Gidney}},
  \citenamefont {{Giustina}}, \citenamefont {{Graff}}, \citenamefont
  {{Guerin}}, \citenamefont {{Habegger}}, \citenamefont {{Harrigan}},
  \citenamefont {{Hartmann}}, \citenamefont {{Ho}}, \citenamefont {{Hoffmann}},
  \citenamefont {{Huang}}, \citenamefont {{Humble}}, \citenamefont {{Isakov}},
  \citenamefont {{Jeffrey}}, \citenamefont {{Jiang}}, \citenamefont {{Kafri}},
  \citenamefont {{Kechedzhi}}, \citenamefont {{Kelly}}, \citenamefont
  {{Klimov}}, \citenamefont {{Knysh}}, \citenamefont {{Korotkov}},
  \citenamefont {{Kostritsa}}, \citenamefont {{Landhuis}}, \citenamefont
  {{Lindmark}}, \citenamefont {{Lucero}}, \citenamefont {{Lyakh}},
  \citenamefont {{Mandr{\`a}}}, \citenamefont {{McClean}}, \citenamefont
  {{McEwen}}, \citenamefont {{Megrant}}, \citenamefont {{Mi}}, \citenamefont
  {{Michielsen}}, \citenamefont {{Mohseni}}, \citenamefont {{Mutus}},
  \citenamefont {{Naaman}}, \citenamefont {{Neeley}}, \citenamefont {{Neill}},
  \citenamefont {{Niu}}, \citenamefont {{Ostby}}, \citenamefont {{Petukhov}},
  \citenamefont {{Platt}}, \citenamefont {{Quintana}}, \citenamefont
  {{Rieffel}}, \citenamefont {{Roushan}}, \citenamefont {{Rubin}},
  \citenamefont {{Sank}}, \citenamefont {{Satzinger}}, \citenamefont
  {{Smelyanskiy}}, \citenamefont {{Sung}}, \citenamefont {{Trevithick}},
  \citenamefont {{Vainsencher}}, \citenamefont {{Villalonga}}, \citenamefont
  {{White}}, \citenamefont {{Yao}}, \citenamefont {{Yeh}}, \citenamefont
  {{Zalcman}}, \citenamefont {{Neven}},\ and\ \citenamefont
  {{Martinis}}}]{2019Nature}%
  \BibitemOpen
  \bibfield  {author} {\bibinfo {author} {\bibfnamefont {Frank}\ \bibnamefont
  {{Arute}}}, \bibinfo {author} {\bibfnamefont {Kunal}\ \bibnamefont {{Arya}}},
  \bibinfo {author} {\bibfnamefont {Ryan}\ \bibnamefont {{Babbush}}}, \bibinfo
  {author} {\bibfnamefont {Dave}\ \bibnamefont {{Bacon}}}, \bibinfo {author}
  {\bibfnamefont {Joseph~C.}\ \bibnamefont {{Bardin}}}, \bibinfo {author}
  {\bibfnamefont {Rami}\ \bibnamefont {{Barends}}}, \bibinfo {author}
  {\bibfnamefont {Rupak}\ \bibnamefont {{Biswas}}}, \bibinfo {author}
  {\bibfnamefont {Sergio}\ \bibnamefont {{Boixo}}}, \bibinfo {author}
  {\bibfnamefont {Fernando G.~S.~L.}\ \bibnamefont {{Brandao}}}, \bibinfo
  {author} {\bibfnamefont {David~A.}\ \bibnamefont {{Buell}}}, \bibinfo
  {author} {\bibfnamefont {Brian}\ \bibnamefont {{Burkett}}}, \bibinfo {author}
  {\bibfnamefont {Yu}~\bibnamefont {{Chen}}}, \bibinfo {author} {\bibfnamefont
  {Zijun}\ \bibnamefont {{Chen}}}, \bibinfo {author} {\bibfnamefont {Ben}\
  \bibnamefont {{Chiaro}}}, \bibinfo {author} {\bibfnamefont {Roberto}\
  \bibnamefont {{Collins}}}, \bibinfo {author} {\bibfnamefont {William}\
  \bibnamefont {{Courtney}}}, \bibinfo {author} {\bibfnamefont {Andrew}\
  \bibnamefont {{Dunsworth}}}, \bibinfo {author} {\bibfnamefont {Edward}\
  \bibnamefont {{Farhi}}}, \bibinfo {author} {\bibfnamefont {Brooks}\
  \bibnamefont {{Foxen}}}, \bibinfo {author} {\bibfnamefont {Austin}\
  \bibnamefont {{Fowler}}}, \bibinfo {author} {\bibfnamefont {Craig}\
  \bibnamefont {{Gidney}}}, \bibinfo {author} {\bibfnamefont {Marissa}\
  \bibnamefont {{Giustina}}}, \bibinfo {author} {\bibfnamefont {Rob}\
  \bibnamefont {{Graff}}}, \bibinfo {author} {\bibfnamefont {Keith}\
  \bibnamefont {{Guerin}}}, \bibinfo {author} {\bibfnamefont {Steve}\
  \bibnamefont {{Habegger}}}, \bibinfo {author} {\bibfnamefont {Matthew~P.}\
  \bibnamefont {{Harrigan}}}, \bibinfo {author} {\bibfnamefont {Michael~J.}\
  \bibnamefont {{Hartmann}}}, \bibinfo {author} {\bibfnamefont {Alan}\
  \bibnamefont {{Ho}}}, \bibinfo {author} {\bibfnamefont {Markus}\ \bibnamefont
  {{Hoffmann}}}, \bibinfo {author} {\bibfnamefont {Trent}\ \bibnamefont
  {{Huang}}}, \bibinfo {author} {\bibfnamefont {Travis~S.}\ \bibnamefont
  {{Humble}}}, \bibinfo {author} {\bibfnamefont {Sergei~V.}\ \bibnamefont
  {{Isakov}}}, \bibinfo {author} {\bibfnamefont {Evan}\ \bibnamefont
  {{Jeffrey}}}, \bibinfo {author} {\bibfnamefont {Zhang}\ \bibnamefont
  {{Jiang}}}, \bibinfo {author} {\bibfnamefont {Dvir}\ \bibnamefont {{Kafri}}},
  \bibinfo {author} {\bibfnamefont {Kostyantyn}\ \bibnamefont {{Kechedzhi}}},
  \bibinfo {author} {\bibfnamefont {Julian}\ \bibnamefont {{Kelly}}}, \bibinfo
  {author} {\bibfnamefont {Paul~V.}\ \bibnamefont {{Klimov}}}, \bibinfo
  {author} {\bibfnamefont {Sergey}\ \bibnamefont {{Knysh}}}, \bibinfo {author}
  {\bibfnamefont {Alexander}\ \bibnamefont {{Korotkov}}}, \bibinfo {author}
  {\bibfnamefont {Fedor}\ \bibnamefont {{Kostritsa}}}, \bibinfo {author}
  {\bibfnamefont {David}\ \bibnamefont {{Landhuis}}}, \bibinfo {author}
  {\bibfnamefont {Mike}\ \bibnamefont {{Lindmark}}}, \bibinfo {author}
  {\bibfnamefont {Erik}\ \bibnamefont {{Lucero}}}, \bibinfo {author}
  {\bibfnamefont {Dmitry}\ \bibnamefont {{Lyakh}}}, \bibinfo {author}
  {\bibfnamefont {Salvatore}\ \bibnamefont {{Mandr{\`a}}}}, \bibinfo {author}
  {\bibfnamefont {Jarrod~R.}\ \bibnamefont {{McClean}}}, \bibinfo {author}
  {\bibfnamefont {Matthew}\ \bibnamefont {{McEwen}}}, \bibinfo {author}
  {\bibfnamefont {Anthony}\ \bibnamefont {{Megrant}}}, \bibinfo {author}
  {\bibfnamefont {Xiao}\ \bibnamefont {{Mi}}}, \bibinfo {author} {\bibfnamefont
  {Kristel}\ \bibnamefont {{Michielsen}}}, \bibinfo {author} {\bibfnamefont
  {Masoud}\ \bibnamefont {{Mohseni}}}, \bibinfo {author} {\bibfnamefont {Josh}\
  \bibnamefont {{Mutus}}}, \bibinfo {author} {\bibfnamefont {Ofer}\
  \bibnamefont {{Naaman}}}, \bibinfo {author} {\bibfnamefont {Matthew}\
  \bibnamefont {{Neeley}}}, \bibinfo {author} {\bibfnamefont {Charles}\
  \bibnamefont {{Neill}}}, \bibinfo {author} {\bibfnamefont {Murphy~Yuezhen}\
  \bibnamefont {{Niu}}}, \bibinfo {author} {\bibfnamefont {Eric}\ \bibnamefont
  {{Ostby}}}, \bibinfo {author} {\bibfnamefont {Andre}\ \bibnamefont
  {{Petukhov}}}, \bibinfo {author} {\bibfnamefont {John~C.}\ \bibnamefont
  {{Platt}}}, \bibinfo {author} {\bibfnamefont {Chris}\ \bibnamefont
  {{Quintana}}}, \bibinfo {author} {\bibfnamefont {Eleanor~G.}\ \bibnamefont
  {{Rieffel}}}, \bibinfo {author} {\bibfnamefont {Pedram}\ \bibnamefont
  {{Roushan}}}, \bibinfo {author} {\bibfnamefont {Nicholas~C.}\ \bibnamefont
  {{Rubin}}}, \bibinfo {author} {\bibfnamefont {Daniel}\ \bibnamefont
  {{Sank}}}, \bibinfo {author} {\bibfnamefont {Kevin~J.}\ \bibnamefont
  {{Satzinger}}}, \bibinfo {author} {\bibfnamefont {Vadim}\ \bibnamefont
  {{Smelyanskiy}}}, \bibinfo {author} {\bibfnamefont {Kevin~J.}\ \bibnamefont
  {{Sung}}}, \bibinfo {author} {\bibfnamefont {Matthew~D.}\ \bibnamefont
  {{Trevithick}}}, \bibinfo {author} {\bibfnamefont {Amit}\ \bibnamefont
  {{Vainsencher}}}, \bibinfo {author} {\bibfnamefont {Benjamin}\ \bibnamefont
  {{Villalonga}}}, \bibinfo {author} {\bibfnamefont {Theodore}\ \bibnamefont
  {{White}}}, \bibinfo {author} {\bibfnamefont {Z.~Jamie}\ \bibnamefont
  {{Yao}}}, \bibinfo {author} {\bibfnamefont {Ping}\ \bibnamefont {{Yeh}}},
  \bibinfo {author} {\bibfnamefont {Adam}\ \bibnamefont {{Zalcman}}}, \bibinfo
  {author} {\bibfnamefont {Hartmut}\ \bibnamefont {{Neven}}}, \ and\ \bibinfo
  {author} {\bibfnamefont {John~M.}\ \bibnamefont {{Martinis}}},\ }\bibfield
  {title} {\enquote {\bibinfo {title} {{Quantum supremacy using a programmable
  superconducting processor}},}\ }\href {\doibase 10.1038/s41586-019-1666-5}
  {\bibfield  {journal} {\bibinfo  {journal} {\nat}\ }\textbf {\bibinfo
  {volume} {574}},\ \bibinfo {pages} {505--510} (\bibinfo {year} {2019})},\
  \Eprint {http://arxiv.org/abs/1910.11333} {arXiv:1910.11333 [quant-ph]}
  \BibitemShut {NoStop}%
\bibitem [{\citenamefont {{Skinner}}\ \emph {et~al.}(2019)\citenamefont
  {{Skinner}}, \citenamefont {{Ruhman}},\ and\ \citenamefont
  {{Nahum}}}]{2019mipt1}%
  \BibitemOpen
  \bibfield  {author} {\bibinfo {author} {\bibfnamefont {Brian}\ \bibnamefont
  {{Skinner}}}, \bibinfo {author} {\bibfnamefont {Jonathan}\ \bibnamefont
  {{Ruhman}}}, \ and\ \bibinfo {author} {\bibfnamefont {Adam}\ \bibnamefont
  {{Nahum}}},\ }\bibfield  {title} {\enquote {\bibinfo {title}
  {{Measurement-Induced Phase Transitions in the Dynamics of Entanglement}},}\
  }\href {\doibase 10.1103/PhysRevX.9.031009} {\bibfield  {journal} {\bibinfo
  {journal} {Physical Review X}\ }\textbf {\bibinfo {volume} {9}},\ \bibinfo
  {eid} {031009} (\bibinfo {year} {2019})},\ \Eprint
  {http://arxiv.org/abs/1808.05953} {arXiv:1808.05953 [cond-mat.stat-mech]}
  \BibitemShut {NoStop}%
\bibitem [{\citenamefont {{Li}}\ \emph {et~al.}(2018)\citenamefont {{Li}},
  \citenamefont {{Chen}},\ and\ \citenamefont {{Fisher}}}]{2018mipt2}%
  \BibitemOpen
  \bibfield  {author} {\bibinfo {author} {\bibfnamefont {Yaodong}\ \bibnamefont
  {{Li}}}, \bibinfo {author} {\bibfnamefont {Xiao}\ \bibnamefont {{Chen}}}, \
  and\ \bibinfo {author} {\bibfnamefont {Matthew P.~A.}\ \bibnamefont
  {{Fisher}}},\ }\bibfield  {title} {\enquote {\bibinfo {title} {{Quantum Zeno
  effect and the many-body entanglement transition}},}\ }\href {\doibase
  10.1103/PhysRevB.98.205136} {\bibfield  {journal} {\bibinfo  {journal}
  {\prb}\ }\textbf {\bibinfo {volume} {98}},\ \bibinfo {eid} {205136} (\bibinfo
  {year} {2018})},\ \Eprint {http://arxiv.org/abs/1808.06134} {arXiv:1808.06134
  [quant-ph]} \BibitemShut {NoStop}%
\bibitem [{\citenamefont {{Chan}}\ \emph {et~al.}(2019)\citenamefont {{Chan}},
  \citenamefont {{Nandkishore}}, \citenamefont {{Pretko}},\ and\ \citenamefont
  {{Smith}}}]{2019chan}%
  \BibitemOpen
  \bibfield  {author} {\bibinfo {author} {\bibfnamefont {Amos}\ \bibnamefont
  {{Chan}}}, \bibinfo {author} {\bibfnamefont {Rahul~M.}\ \bibnamefont
  {{Nandkishore}}}, \bibinfo {author} {\bibfnamefont {Michael}\ \bibnamefont
  {{Pretko}}}, \ and\ \bibinfo {author} {\bibfnamefont {Graeme}\ \bibnamefont
  {{Smith}}},\ }\bibfield  {title} {\enquote {\bibinfo {title}
  {{Unitary-projective entanglement dynamics}},}\ }\href {\doibase
  10.1103/PhysRevB.99.224307} {\bibfield  {journal} {\bibinfo  {journal}
  {\prb}\ }\textbf {\bibinfo {volume} {99}},\ \bibinfo {eid} {224307} (\bibinfo
  {year} {2019})},\ \Eprint {http://arxiv.org/abs/1808.05949} {arXiv:1808.05949
  [cond-mat.stat-mech]} \BibitemShut {NoStop}%
\bibitem [{\citenamefont {{Gullans}}\ and\ \citenamefont
  {{Huse}}(2020)}]{2020gullans}%
  \BibitemOpen
  \bibfield  {author} {\bibinfo {author} {\bibfnamefont {Michael~J.}\
  \bibnamefont {{Gullans}}}\ and\ \bibinfo {author} {\bibfnamefont {David~A.}\
  \bibnamefont {{Huse}}},\ }\bibfield  {title} {\enquote {\bibinfo {title}
  {{Dynamical Purification Phase Transition Induced by Quantum
  Measurements}},}\ }\href {\doibase 10.1103/PhysRevX.10.041020} {\bibfield
  {journal} {\bibinfo  {journal} {Physical Review X}\ }\textbf {\bibinfo
  {volume} {10}},\ \bibinfo {eid} {041020} (\bibinfo {year} {2020})},\ \Eprint
  {http://arxiv.org/abs/1905.05195} {arXiv:1905.05195 [quant-ph]} \BibitemShut
  {NoStop}%
\bibitem [{\citenamefont {{Jian}}\ \emph
  {et~al.}(2020{\natexlab{a}})\citenamefont {{Jian}}, \citenamefont {{You}},
  \citenamefont {{Vasseur}},\ and\ \citenamefont {{Ludwig}}}]{2020jianyou}%
  \BibitemOpen
  \bibfield  {author} {\bibinfo {author} {\bibfnamefont {Chao-Ming}\
  \bibnamefont {{Jian}}}, \bibinfo {author} {\bibfnamefont {Yi-Zhuang}\
  \bibnamefont {{You}}}, \bibinfo {author} {\bibfnamefont {Romain}\
  \bibnamefont {{Vasseur}}}, \ and\ \bibinfo {author} {\bibfnamefont {Andreas
  W.~W.}\ \bibnamefont {{Ludwig}}},\ }\bibfield  {title} {\enquote {\bibinfo
  {title} {{Measurement-induced criticality in random quantum circuits}},}\
  }\href {\doibase 10.1103/PhysRevB.101.104302} {\bibfield  {journal} {\bibinfo
   {journal} {\prb}\ }\textbf {\bibinfo {volume} {101}},\ \bibinfo {eid}
  {104302} (\bibinfo {year} {2020}{\natexlab{a}})},\ \Eprint
  {http://arxiv.org/abs/1908.08051} {arXiv:1908.08051 [cond-mat.stat-mech]}
  \BibitemShut {NoStop}%
\bibitem [{\citenamefont {{Bao}}\ \emph {et~al.}(2020)\citenamefont {{Bao}},
  \citenamefont {{Choi}},\ and\ \citenamefont {{Altman}}}]{2020baochoi}%
  \BibitemOpen
  \bibfield  {author} {\bibinfo {author} {\bibfnamefont {Yimu}\ \bibnamefont
  {{Bao}}}, \bibinfo {author} {\bibfnamefont {Soonwon}\ \bibnamefont {{Choi}}},
  \ and\ \bibinfo {author} {\bibfnamefont {Ehud}\ \bibnamefont {{Altman}}},\
  }\bibfield  {title} {\enquote {\bibinfo {title} {{Theory of the phase
  transition in random unitary circuits with measurements}},}\ }\href {\doibase
  10.1103/PhysRevB.101.104301} {\bibfield  {journal} {\bibinfo  {journal}
  {\prb}\ }\textbf {\bibinfo {volume} {101}},\ \bibinfo {eid} {104301}
  (\bibinfo {year} {2020})},\ \Eprint {http://arxiv.org/abs/1908.04305}
  {arXiv:1908.04305 [cond-mat.stat-mech]} \BibitemShut {NoStop}%
\bibitem [{\citenamefont {{Bao}}\ \emph {et~al.}(2023)\citenamefont {{Bao}},
  \citenamefont {{Fan}}, \citenamefont {{Vishwanath}},\ and\ \citenamefont
  {{Altman}}}]{2023bao}%
  \BibitemOpen
  \bibfield  {author} {\bibinfo {author} {\bibfnamefont {Yimu}\ \bibnamefont
  {{Bao}}}, \bibinfo {author} {\bibfnamefont {Ruihua}\ \bibnamefont {{Fan}}},
  \bibinfo {author} {\bibfnamefont {Ashvin}\ \bibnamefont {{Vishwanath}}}, \
  and\ \bibinfo {author} {\bibfnamefont {Ehud}\ \bibnamefont {{Altman}}},\
  }\bibfield  {title} {\enquote {\bibinfo {title} {{Mixed-state topological
  order and the errorfield double formulation of decoherence-induced
  transitions}},}\ }\href {\doibase 10.48550/arXiv.2301.05687} {\bibfield
  {journal} {\bibinfo  {journal} {arXiv e-prints}\ ,\ \bibinfo {eid}
  {arXiv:2301.05687}} (\bibinfo {year} {2023})},\ \Eprint
  {http://arxiv.org/abs/2301.05687} {arXiv:2301.05687 [quant-ph]} \BibitemShut
  {NoStop}%
\bibitem [{\citenamefont {{Fan}}\ \emph {et~al.}(2023)\citenamefont {{Fan}},
  \citenamefont {{Bao}}, \citenamefont {{Altman}},\ and\ \citenamefont
  {{Vishwanath}}}]{2023Fan}%
  \BibitemOpen
  \bibfield  {author} {\bibinfo {author} {\bibfnamefont {Ruihua}\ \bibnamefont
  {{Fan}}}, \bibinfo {author} {\bibfnamefont {Yimu}\ \bibnamefont {{Bao}}},
  \bibinfo {author} {\bibfnamefont {Ehud}\ \bibnamefont {{Altman}}}, \ and\
  \bibinfo {author} {\bibfnamefont {Ashvin}\ \bibnamefont {{Vishwanath}}},\
  }\bibfield  {title} {\enquote {\bibinfo {title} {{Diagnostics of mixed-state
  topological order and breakdown of quantum memory}},}\ }\href {\doibase
  10.48550/arXiv.2301.05689} {\bibfield  {journal} {\bibinfo  {journal} {arXiv
  e-prints}\ ,\ \bibinfo {eid} {arXiv:2301.05689}} (\bibinfo {year} {2023})},\
  \Eprint {http://arxiv.org/abs/2301.05689} {arXiv:2301.05689 [quant-ph]}
  \BibitemShut {NoStop}%
\bibitem [{\citenamefont {{Lee}}\ \emph
  {et~al.}(2023{\natexlab{a}})\citenamefont {{Lee}}, \citenamefont {{Jian}},\
  and\ \citenamefont {{Xu}}}]{2023decoxu}%
  \BibitemOpen
  \bibfield  {author} {\bibinfo {author} {\bibfnamefont {Jong~Yeon}\
  \bibnamefont {{Lee}}}, \bibinfo {author} {\bibfnamefont {Chao-Ming}\
  \bibnamefont {{Jian}}}, \ and\ \bibinfo {author} {\bibfnamefont {Cenke}\
  \bibnamefont {{Xu}}},\ }\bibfield  {title} {\enquote {\bibinfo {title}
  {{Quantum criticality under decoherence or weak measurement}},}\ }\href
  {\doibase 10.48550/arXiv.2301.05238} {\bibfield  {journal} {\bibinfo
  {journal} {arXiv e-prints}\ ,\ \bibinfo {eid} {arXiv:2301.05238}} (\bibinfo
  {year} {2023}{\natexlab{a}})},\ \Eprint {http://arxiv.org/abs/2301.05238}
  {arXiv:2301.05238 [cond-mat.stat-mech]} \BibitemShut {NoStop}%
\bibitem [{\citenamefont {{Zou}}\ \emph {et~al.}(2023)\citenamefont {{Zou}},
  \citenamefont {{Sang}},\ and\ \citenamefont {{Hsieh}}}]{2023tim}%
  \BibitemOpen
  \bibfield  {author} {\bibinfo {author} {\bibfnamefont {Yijian}\ \bibnamefont
  {{Zou}}}, \bibinfo {author} {\bibfnamefont {Shengqi}\ \bibnamefont {{Sang}}},
  \ and\ \bibinfo {author} {\bibfnamefont {Timothy~H.}\ \bibnamefont
  {{Hsieh}}},\ }\bibfield  {title} {\enquote {\bibinfo {title} {{Channeling
  quantum criticality}},}\ }\href {\doibase 10.48550/arXiv.2301.07141}
  {\bibfield  {journal} {\bibinfo  {journal} {arXiv e-prints}\ ,\ \bibinfo
  {eid} {arXiv:2301.07141}} (\bibinfo {year} {2023})},\ \Eprint
  {http://arxiv.org/abs/2301.07141} {arXiv:2301.07141 [quant-ph]} \BibitemShut
  {NoStop}%
\bibitem [{\citenamefont {{Sieberer}}\ \emph {et~al.}(2016)\citenamefont
  {{Sieberer}}, \citenamefont {{Buchhold}},\ and\ \citenamefont
  {{Diehl}}}]{2016Diehl}%
  \BibitemOpen
  \bibfield  {author} {\bibinfo {author} {\bibfnamefont {L.~M.}\ \bibnamefont
  {{Sieberer}}}, \bibinfo {author} {\bibfnamefont {M.}~\bibnamefont
  {{Buchhold}}}, \ and\ \bibinfo {author} {\bibfnamefont {S.}~\bibnamefont
  {{Diehl}}},\ }\bibfield  {title} {\enquote {\bibinfo {title} {{Keldysh field
  theory for driven open quantum systems}},}\ }\href {\doibase
  10.1088/0034-4885/79/9/096001} {\bibfield  {journal} {\bibinfo  {journal}
  {Reports on Progress in Physics}\ }\textbf {\bibinfo {volume} {79}},\
  \bibinfo {eid} {096001} (\bibinfo {year} {2016})},\ \Eprint
  {http://arxiv.org/abs/1512.00637} {arXiv:1512.00637 [cond-mat.quant-gas]}
  \BibitemShut {NoStop}%
\bibitem [{\citenamefont {Kamenev}(2023)}]{kamenev2023field}%
  \BibitemOpen
  \bibfield  {author} {\bibinfo {author} {\bibfnamefont {Alex}\ \bibnamefont
  {Kamenev}},\ }\href@noop {} {\emph {\bibinfo {title} {Field theory of
  non-equilibrium systems}}}\ (\bibinfo  {publisher} {Cambridge University
  Press},\ \bibinfo {year} {2023})\BibitemShut {NoStop}%
\bibitem [{\citenamefont {{Weinstein}}\ \emph {et~al.}(2023)\citenamefont
  {{Weinstein}}, \citenamefont {{Sajith}}, \citenamefont {{Altman}},\ and\
  \citenamefont {{Garratt}}}]{2023ehud}%
  \BibitemOpen
  \bibfield  {author} {\bibinfo {author} {\bibfnamefont {Zack}\ \bibnamefont
  {{Weinstein}}}, \bibinfo {author} {\bibfnamefont {Rohith}\ \bibnamefont
  {{Sajith}}}, \bibinfo {author} {\bibfnamefont {Ehud}\ \bibnamefont
  {{Altman}}}, \ and\ \bibinfo {author} {\bibfnamefont {Samuel~J.}\
  \bibnamefont {{Garratt}}},\ }\bibfield  {title} {\enquote {\bibinfo {title}
  {{Nonlocality and entanglement in measured critical quantum Ising chains}},}\
  }\href {\doibase 10.48550/arXiv.2301.08268} {\bibfield  {journal} {\bibinfo
  {journal} {arXiv e-prints}\ ,\ \bibinfo {eid} {arXiv:2301.08268}} (\bibinfo
  {year} {2023})},\ \Eprint {http://arxiv.org/abs/2301.08268} {arXiv:2301.08268
  [cond-mat.stat-mech]} \BibitemShut {NoStop}%
\bibitem [{\citenamefont {{Lin}}\ \emph {et~al.}(2023)\citenamefont {{Lin}},
  \citenamefont {{Ye}}, \citenamefont {{Zou}}, \citenamefont {{Sang}},\ and\
  \citenamefont {{Hsieh}}}]{2023linC}%
  \BibitemOpen
  \bibfield  {author} {\bibinfo {author} {\bibfnamefont {Cheng-Ju}\
  \bibnamefont {{Lin}}}, \bibinfo {author} {\bibfnamefont {Weicheng}\
  \bibnamefont {{Ye}}}, \bibinfo {author} {\bibfnamefont {Yijian}\ \bibnamefont
  {{Zou}}}, \bibinfo {author} {\bibfnamefont {Shengqi}\ \bibnamefont {{Sang}}},
  \ and\ \bibinfo {author} {\bibfnamefont {Timothy~H.}\ \bibnamefont
  {{Hsieh}}},\ }\bibfield  {title} {\enquote {\bibinfo {title} {{Probing sign
  structure using measurement-induced entanglement}},}\ }\href {\doibase
  10.22331/q-2023-02-02-910} {\bibfield  {journal} {\bibinfo  {journal}
  {Quantum}\ }\textbf {\bibinfo {volume} {7}},\ \bibinfo {pages} {910}
  (\bibinfo {year} {2023})},\ \Eprint {http://arxiv.org/abs/2205.05692}
  {arXiv:2205.05692 [quant-ph]} \BibitemShut {NoStop}%
\bibitem [{\citenamefont {{Garratt}}\ \emph {et~al.}(2022)\citenamefont
  {{Garratt}}, \citenamefont {{Weinstein}},\ and\ \citenamefont
  {{Altman}}}]{2022altman}%
  \BibitemOpen
  \bibfield  {author} {\bibinfo {author} {\bibfnamefont {Samuel~J.}\
  \bibnamefont {{Garratt}}}, \bibinfo {author} {\bibfnamefont {Zack}\
  \bibnamefont {{Weinstein}}}, \ and\ \bibinfo {author} {\bibfnamefont {Ehud}\
  \bibnamefont {{Altman}}},\ }\bibfield  {title} {\enquote {\bibinfo {title}
  {{Measurements conspire nonlocally to restructure critical quantum
  states}},}\ }\href {\doibase 10.48550/arXiv.2207.09476} {\bibfield  {journal}
  {\bibinfo  {journal} {arXiv e-prints}\ ,\ \bibinfo {eid} {arXiv:2207.09476}}
  (\bibinfo {year} {2022})},\ \Eprint {http://arxiv.org/abs/2207.09476}
  {arXiv:2207.09476 [cond-mat.stat-mech]} \BibitemShut {NoStop}%
\bibitem [{\citenamefont {{Yang}}\ \emph {et~al.}(2023)\citenamefont {{Yang}},
  \citenamefont {{Mao}},\ and\ \citenamefont {{Jian}}}]{2023jian}%
  \BibitemOpen
  \bibfield  {author} {\bibinfo {author} {\bibfnamefont {Zhou}\ \bibnamefont
  {{Yang}}}, \bibinfo {author} {\bibfnamefont {Dan}\ \bibnamefont {{Mao}}}, \
  and\ \bibinfo {author} {\bibfnamefont {Chao-Ming}\ \bibnamefont {{Jian}}},\
  }\bibfield  {title} {\enquote {\bibinfo {title} {{Entanglement in
  one-dimensional critical state after measurements}},}\ }\href {\doibase
  10.48550/arXiv.2301.08255} {\bibfield  {journal} {\bibinfo  {journal} {arXiv
  e-prints}\ ,\ \bibinfo {eid} {arXiv:2301.08255}} (\bibinfo {year} {2023})},\
  \Eprint {http://arxiv.org/abs/2301.08255} {arXiv:2301.08255 [quant-ph]}
  \BibitemShut {NoStop}%
\bibitem [{\citenamefont {{Jacobs}}\ and\ \citenamefont
  {{Steck}}(2006{\natexlab{a}})}]{2006QSDJ}%
  \BibitemOpen
  \bibfield  {author} {\bibinfo {author} {\bibfnamefont {Kurt}\ \bibnamefont
  {{Jacobs}}}\ and\ \bibinfo {author} {\bibfnamefont {Daniel}\ \bibnamefont
  {{Steck}}},\ }\bibfield  {title} {\enquote {\bibinfo {title} {{A
  straightforward introduction to continuous quantum measurement}},}\ }\href
  {\doibase 10.1080/00107510601101934} {\bibfield  {journal} {\bibinfo
  {journal} {Contemporary Physics}\ }\textbf {\bibinfo {volume} {47}},\
  \bibinfo {pages} {279--303} (\bibinfo {year} {2006}{\natexlab{a}})},\ \Eprint
  {http://arxiv.org/abs/quant-ph/0611067} {arXiv:quant-ph/0611067 [quant-ph]}
  \BibitemShut {NoStop}%
\bibitem [{\citenamefont {{Lee}}\ \emph
  {et~al.}(2023{\natexlab{b}})\citenamefont {{Lee}}, \citenamefont {{Jian}},\
  and\ \citenamefont {{Xu}}}]{2023xu}%
  \BibitemOpen
  \bibfield  {author} {\bibinfo {author} {\bibfnamefont {Jong~Yeon}\
  \bibnamefont {{Lee}}}, \bibinfo {author} {\bibfnamefont {Chao-Ming}\
  \bibnamefont {{Jian}}}, \ and\ \bibinfo {author} {\bibfnamefont {Cenke}\
  \bibnamefont {{Xu}}},\ }\bibfield  {title} {\enquote {\bibinfo {title}
  {{Quantum criticality under decoherence or weak measurement}},}\ }\href
  {\doibase 10.48550/arXiv.2301.05238} {\bibfield  {journal} {\bibinfo
  {journal} {arXiv e-prints}\ ,\ \bibinfo {eid} {arXiv:2301.05238}} (\bibinfo
  {year} {2023}{\natexlab{b}})},\ \Eprint {http://arxiv.org/abs/2301.05238}
  {arXiv:2301.05238 [cond-mat.stat-mech]} \BibitemShut {NoStop}%
\bibitem [{\citenamefont {{Buchhold}}\ \emph {et~al.}(2021)\citenamefont
  {{Buchhold}}, \citenamefont {{Minoguchi}}, \citenamefont {{Altland}},\ and\
  \citenamefont {{Diehl}}}]{2021buchhold}%
  \BibitemOpen
  \bibfield  {author} {\bibinfo {author} {\bibfnamefont {M.}~\bibnamefont
  {{Buchhold}}}, \bibinfo {author} {\bibfnamefont {Y.}~\bibnamefont
  {{Minoguchi}}}, \bibinfo {author} {\bibfnamefont {A.}~\bibnamefont
  {{Altland}}}, \ and\ \bibinfo {author} {\bibfnamefont {S.}~\bibnamefont
  {{Diehl}}},\ }\bibfield  {title} {\enquote {\bibinfo {title} {{Effective
  Theory for the Measurement-Induced Phase Transition of Dirac Fermions}},}\
  }\href {\doibase 10.1103/PhysRevX.11.041004} {\bibfield  {journal} {\bibinfo
  {journal} {Physical Review X}\ }\textbf {\bibinfo {volume} {11}},\ \bibinfo
  {eid} {041004} (\bibinfo {year} {2021})},\ \Eprint
  {http://arxiv.org/abs/2102.08381} {arXiv:2102.08381 [cond-mat.stat-mech]}
  \BibitemShut {NoStop}%
\bibitem [{\citenamefont {{Ladewig}}\ \emph {et~al.}(2022)\citenamefont
  {{Ladewig}}, \citenamefont {{Diehl}},\ and\ \citenamefont
  {{Buchhold}}}]{2022ladewig}%
  \BibitemOpen
  \bibfield  {author} {\bibinfo {author} {\bibfnamefont {B.}~\bibnamefont
  {{Ladewig}}}, \bibinfo {author} {\bibfnamefont {S.}~\bibnamefont {{Diehl}}},
  \ and\ \bibinfo {author} {\bibfnamefont {M.}~\bibnamefont {{Buchhold}}},\
  }\bibfield  {title} {\enquote {\bibinfo {title} {{Monitored open fermion
  dynamics: Exploring the interplay of measurement, decoherence, and free
  Hamiltonian evolution}},}\ }\href {\doibase 10.1103/PhysRevResearch.4.033001}
  {\bibfield  {journal} {\bibinfo  {journal} {Physical Review Research}\
  }\textbf {\bibinfo {volume} {4}},\ \bibinfo {eid} {033001} (\bibinfo {year}
  {2022})},\ \Eprint {http://arxiv.org/abs/2203.00027} {arXiv:2203.00027
  [cond-mat.stat-mech]} \BibitemShut {NoStop}%
\bibitem [{\citenamefont {{Kamenev}}\ and\ \citenamefont
  {{Levchenko}}(2009)}]{2009Kamenev}%
  \BibitemOpen
  \bibfield  {author} {\bibinfo {author} {\bibfnamefont {Alex}\ \bibnamefont
  {{Kamenev}}}\ and\ \bibinfo {author} {\bibfnamefont {Alex}\ \bibnamefont
  {{Levchenko}}},\ }\bibfield  {title} {\enquote {\bibinfo {title} {{Keldysh
  technique and non-linear {\ensuremath{\sigma}}-model: basic principles and
  applications}},}\ }\href {\doibase 10.1080/00018730902850504} {\bibfield
  {journal} {\bibinfo  {journal} {Advances in Physics}\ }\textbf {\bibinfo
  {volume} {58}},\ \bibinfo {pages} {197--319} (\bibinfo {year} {2009})},\
  \Eprint {http://arxiv.org/abs/0901.3586} {arXiv:0901.3586 [cond-mat.other]}
  \BibitemShut {NoStop}%
\bibitem [{\citenamefont {{Bu{\v{c}}a}}\ and\ \citenamefont
  {{Prosen}}(2012)}]{2012strongweak}%
  \BibitemOpen
  \bibfield  {author} {\bibinfo {author} {\bibfnamefont {Berislav}\
  \bibnamefont {{Bu{\v{c}}a}}}\ and\ \bibinfo {author} {\bibfnamefont
  {Toma{\v{z}}}\ \bibnamefont {{Prosen}}},\ }\bibfield  {title} {\enquote
  {\bibinfo {title} {{A note on symmetry reductions of the Lindblad equation:
  transport in constrained open spin chains}},}\ }\href {\doibase
  10.1088/1367-2630/14/7/073007} {\bibfield  {journal} {\bibinfo  {journal}
  {New Journal of Physics}\ }\textbf {\bibinfo {volume} {14}},\ \bibinfo {eid}
  {073007} (\bibinfo {year} {2012})},\ \Eprint {http://arxiv.org/abs/1203.0943}
  {arXiv:1203.0943 [quant-ph]} \BibitemShut {NoStop}%
\bibitem [{\citenamefont {{de Groot}}\ \emph {et~al.}(2022)\citenamefont {{de
  Groot}}, \citenamefont {{Turzillo}},\ and\ \citenamefont
  {{Schuch}}}]{2022turzillo}%
  \BibitemOpen
  \bibfield  {author} {\bibinfo {author} {\bibfnamefont {Caroline}\
  \bibnamefont {{de Groot}}}, \bibinfo {author} {\bibfnamefont {Alex}\
  \bibnamefont {{Turzillo}}}, \ and\ \bibinfo {author} {\bibfnamefont
  {Norbert}\ \bibnamefont {{Schuch}}},\ }\bibfield  {title} {\enquote {\bibinfo
  {title} {{Symmetry Protected Topological Order in Open Quantum Systems}},}\
  }\href {\doibase 10.22331/q-2022-11-10-856} {\bibfield  {journal} {\bibinfo
  {journal} {Quantum}\ }\textbf {\bibinfo {volume} {6}},\ \bibinfo {pages}
  {856} (\bibinfo {year} {2022})},\ \Eprint {http://arxiv.org/abs/2112.04483}
  {arXiv:2112.04483 [quant-ph]} \BibitemShut {NoStop}%
\bibitem [{\citenamefont {Gorini}\ \emph {et~al.}(1976)\citenamefont {Gorini},
  \citenamefont {Kossakowski},\ and\ \citenamefont
  {Sudarshan}}]{gorini1976completely}%
  \BibitemOpen
  \bibfield  {author} {\bibinfo {author} {\bibfnamefont {Vittorio}\
  \bibnamefont {Gorini}}, \bibinfo {author} {\bibfnamefont {Andrzej}\
  \bibnamefont {Kossakowski}}, \ and\ \bibinfo {author} {\bibfnamefont
  {Ennackal Chandy~George}\ \bibnamefont {Sudarshan}},\ }\bibfield  {title}
  {\enquote {\bibinfo {title} {Completely positive dynamical semigroups of
  n-level systems},}\ }\href@noop {} {\bibfield  {journal} {\bibinfo  {journal}
  {Journal of Mathematical Physics}\ }\textbf {\bibinfo {volume} {17}},\
  \bibinfo {pages} {821--825} (\bibinfo {year} {1976})}\BibitemShut {NoStop}%
\bibitem [{\citenamefont {{Jacobs}}\ and\ \citenamefont
  {{Steck}}(2006{\natexlab{b}})}]{2006weakmeasure}%
  \BibitemOpen
  \bibfield  {author} {\bibinfo {author} {\bibfnamefont {Kurt}\ \bibnamefont
  {{Jacobs}}}\ and\ \bibinfo {author} {\bibfnamefont {Daniel}\ \bibnamefont
  {{Steck}}},\ }\bibfield  {title} {\enquote {\bibinfo {title} {{A
  straightforward introduction to continuous quantum measurement}},}\ }\href
  {\doibase 10.1080/00107510601101934} {\bibfield  {journal} {\bibinfo
  {journal} {Contemporary Physics}\ }\textbf {\bibinfo {volume} {47}},\
  \bibinfo {pages} {279--303} (\bibinfo {year} {2006}{\natexlab{b}})},\ \Eprint
  {http://arxiv.org/abs/quant-ph/0611067} {arXiv:quant-ph/0611067 [quant-ph]}
  \BibitemShut {NoStop}%
\bibitem [{\citenamefont {{Brun}}(2002)}]{2002trajectory}%
  \BibitemOpen
  \bibfield  {author} {\bibinfo {author} {\bibfnamefont {Todd~A.}\ \bibnamefont
  {{Brun}}},\ }\bibfield  {title} {\enquote {\bibinfo {title} {{A simple model
  of quantum trajectories}},}\ }\href {\doibase 10.1119/1.1475328} {\bibfield
  {journal} {\bibinfo  {journal} {American Journal of Physics}\ }\textbf
  {\bibinfo {volume} {70}},\ \bibinfo {pages} {719--737} (\bibinfo {year}
  {2002})},\ \Eprint {http://arxiv.org/abs/quant-ph/0108132}
  {arXiv:quant-ph/0108132 [quant-ph]} \BibitemShut {NoStop}%
\bibitem [{\citenamefont {Gisin}\ and\ \citenamefont
  {Percival}(1992)}]{gisin1992quantum}%
  \BibitemOpen
  \bibfield  {author} {\bibinfo {author} {\bibfnamefont {Nicolas}\ \bibnamefont
  {Gisin}}\ and\ \bibinfo {author} {\bibfnamefont {Ian~C}\ \bibnamefont
  {Percival}},\ }\bibfield  {title} {\enquote {\bibinfo {title} {The
  quantum-state diffusion model applied to open systems},}\ }\href@noop {}
  {\bibfield  {journal} {\bibinfo  {journal} {Journal of Physics A:
  Mathematical and General}\ }\textbf {\bibinfo {volume} {25}},\ \bibinfo
  {pages} {5677} (\bibinfo {year} {1992})}\BibitemShut {NoStop}%
\bibitem [{\citenamefont {Wiseman}\ and\ \citenamefont
  {Milburn}(1993)}]{wiseman1993interpretation}%
  \BibitemOpen
  \bibfield  {author} {\bibinfo {author} {\bibfnamefont {HM}~\bibnamefont
  {Wiseman}}\ and\ \bibinfo {author} {\bibfnamefont {GJ}~\bibnamefont
  {Milburn}},\ }\bibfield  {title} {\enquote {\bibinfo {title} {Interpretation
  of quantum jump and diffusion processes illustrated on the bloch sphere},}\
  }\href@noop {} {\bibfield  {journal} {\bibinfo  {journal} {Physical Review
  A}\ }\textbf {\bibinfo {volume} {47}},\ \bibinfo {pages} {1652} (\bibinfo
  {year} {1993})}\BibitemShut {NoStop}%
\bibitem [{\citenamefont {{Jian}}\ \emph
  {et~al.}(2020{\natexlab{b}})\citenamefont {{Jian}}, \citenamefont {{Bauer}},
  \citenamefont {{Keselman}},\ and\ \citenamefont {{Ludwig}}}]{2020Jian}%
  \BibitemOpen
  \bibfield  {author} {\bibinfo {author} {\bibfnamefont {Chao-Ming}\
  \bibnamefont {{Jian}}}, \bibinfo {author} {\bibfnamefont {Bela}\ \bibnamefont
  {{Bauer}}}, \bibinfo {author} {\bibfnamefont {Anna}\ \bibnamefont
  {{Keselman}}}, \ and\ \bibinfo {author} {\bibfnamefont {Andreas W.~W.}\
  \bibnamefont {{Ludwig}}},\ }\bibfield  {title} {\enquote {\bibinfo {title}
  {{Criticality and entanglement in non-unitary quantum circuits and tensor
  networks of non-interacting fermions}},}\ }\href {\doibase
  10.48550/arXiv.2012.04666} {\bibfield  {journal} {\bibinfo  {journal} {arXiv
  e-prints}\ ,\ \bibinfo {eid} {arXiv:2012.04666}} (\bibinfo {year}
  {2020}{\natexlab{b}})},\ \Eprint {http://arxiv.org/abs/2012.04666}
  {arXiv:2012.04666 [cond-mat.stat-mech]} \BibitemShut {NoStop}%
\bibitem [{\citenamefont {{Bao}}\ \emph {et~al.}(2021)\citenamefont {{Bao}},
  \citenamefont {{Choi}},\ and\ \citenamefont {{Altman}}}]{2021Bao}%
  \BibitemOpen
  \bibfield  {author} {\bibinfo {author} {\bibfnamefont {Yimu}\ \bibnamefont
  {{Bao}}}, \bibinfo {author} {\bibfnamefont {Soonwon}\ \bibnamefont {{Choi}}},
  \ and\ \bibinfo {author} {\bibfnamefont {Ehud}\ \bibnamefont {{Altman}}},\
  }\bibfield  {title} {\enquote {\bibinfo {title} {{Symmetry enriched phases of
  quantum circuits}},}\ }\href {\doibase 10.1016/j.aop.2021.168618} {\bibfield
  {journal} {\bibinfo  {journal} {Annals of Physics}\ }\textbf {\bibinfo
  {volume} {435}},\ \bibinfo {eid} {168618} (\bibinfo {year} {2021})},\ \Eprint
  {http://arxiv.org/abs/2102.09164} {arXiv:2102.09164 [cond-mat.stat-mech]}
  \BibitemShut {NoStop}%
\bibitem [{\citenamefont {{Eisler}}\ and\ \citenamefont
  {{Peschel}}(2010)}]{2010defectentro}%
  \BibitemOpen
  \bibfield  {author} {\bibinfo {author} {\bibfnamefont {Viktor}\ \bibnamefont
  {{Eisler}}}\ and\ \bibinfo {author} {\bibfnamefont {Ingo}\ \bibnamefont
  {{Peschel}}},\ }\bibfield  {title} {\enquote {\bibinfo {title} {{Solution of
  the fermionic entanglement problem with interface defects}},}\ }\href
  {\doibase 10.48550/arXiv.1005.2144} {\bibfield  {journal} {\bibinfo
  {journal} {arXiv e-prints}\ ,\ \bibinfo {eid} {arXiv:1005.2144}} (\bibinfo
  {year} {2010})},\ \Eprint {http://arxiv.org/abs/1005.2144} {arXiv:1005.2144
  [cond-mat.stat-mech]} \BibitemShut {NoStop}%
\bibitem [{\citenamefont {{Peschel}}\ and\ \citenamefont
  {{Eisler}}(2012)}]{2012renyidefect}%
  \BibitemOpen
  \bibfield  {author} {\bibinfo {author} {\bibfnamefont {Ingo}\ \bibnamefont
  {{Peschel}}}\ and\ \bibinfo {author} {\bibfnamefont {Viktor}\ \bibnamefont
  {{Eisler}}},\ }\bibfield  {title} {\enquote {\bibinfo {title} {{Exact results
  for the entanglement across defects in critical chains}},}\ }\href {\doibase
  10.1088/1751-8113/45/15/155301} {\bibfield  {journal} {\bibinfo  {journal}
  {Journal of Physics A Mathematical General}\ }\textbf {\bibinfo {volume}
  {45}},\ \bibinfo {eid} {155301} (\bibinfo {year} {2012})},\ \Eprint
  {http://arxiv.org/abs/1201.4104} {arXiv:1201.4104 [cond-mat.stat-mech]}
  \BibitemShut {NoStop}%
\bibitem [{\citenamefont {{Brehm}}\ and\ \citenamefont
  {{Brunner}}(2015)}]{2015defectentro}%
  \BibitemOpen
  \bibfield  {author} {\bibinfo {author} {\bibfnamefont {Enrico~M.}\
  \bibnamefont {{Brehm}}}\ and\ \bibinfo {author} {\bibfnamefont {Ilka}\
  \bibnamefont {{Brunner}}},\ }\bibfield  {title} {\enquote {\bibinfo {title}
  {{Entanglement entropy through conformal interfaces in the 2D Ising
  model}},}\ }\href {\doibase 10.48550/arXiv.1505.02647} {\bibfield  {journal}
  {\bibinfo  {journal} {arXiv e-prints}\ ,\ \bibinfo {eid} {arXiv:1505.02647}}
  (\bibinfo {year} {2015})},\ \Eprint {http://arxiv.org/abs/1505.02647}
  {arXiv:1505.02647 [hep-th]} \BibitemShut {NoStop}%
\bibitem [{\citenamefont {Cardy}(1986)}]{cardy1986effect}%
  \BibitemOpen
  \bibfield  {author} {\bibinfo {author} {\bibfnamefont {John~L}\ \bibnamefont
  {Cardy}},\ }\bibfield  {title} {\enquote {\bibinfo {title} {Effect of
  boundary conditions on the operator content of two-dimensional conformally
  invariant theories},}\ }\href@noop {} {\bibfield  {journal} {\bibinfo
  {journal} {Nuclear Physics B}\ }\textbf {\bibinfo {volume} {275}},\ \bibinfo
  {pages} {200--218} (\bibinfo {year} {1986})}\BibitemShut {NoStop}%
\bibitem [{\citenamefont {Cardy}(1989)}]{cardy1989boundary}%
  \BibitemOpen
  \bibfield  {author} {\bibinfo {author} {\bibfnamefont {John~L}\ \bibnamefont
  {Cardy}},\ }\bibfield  {title} {\enquote {\bibinfo {title} {Boundary
  conditions, fusion rules and the verlinde formula},}\ }\href@noop {}
  {\bibfield  {journal} {\bibinfo  {journal} {Nuclear Physics B}\ }\textbf
  {\bibinfo {volume} {324}},\ \bibinfo {pages} {581--596} (\bibinfo {year}
  {1989})}\BibitemShut {NoStop}%
\bibitem [{\citenamefont {{Calabrese}}\ and\ \citenamefont
  {{Cardy}}(2009)}]{2009cardy}%
  \BibitemOpen
  \bibfield  {author} {\bibinfo {author} {\bibfnamefont {Pasquale}\
  \bibnamefont {{Calabrese}}}\ and\ \bibinfo {author} {\bibfnamefont {John}\
  \bibnamefont {{Cardy}}},\ }\bibfield  {title} {\enquote {\bibinfo {title}
  {{Entanglement entropy and conformal field theory}},}\ }\href {\doibase
  10.1088/1751-8113/42/50/504005} {\bibfield  {journal} {\bibinfo  {journal}
  {Journal of Physics A Mathematical General}\ }\textbf {\bibinfo {volume}
  {42}},\ \bibinfo {eid} {504005} (\bibinfo {year} {2009})},\ \Eprint
  {http://arxiv.org/abs/0905.4013} {arXiv:0905.4013 [cond-mat.stat-mech]}
  \BibitemShut {NoStop}%
\bibitem [{\citenamefont {{Calabrese}}\ and\ \citenamefont
  {{Cardy}}(2005)}]{2005CardyCala}%
  \BibitemOpen
  \bibfield  {author} {\bibinfo {author} {\bibfnamefont {Pasquale}\
  \bibnamefont {{Calabrese}}}\ and\ \bibinfo {author} {\bibfnamefont {John}\
  \bibnamefont {{Cardy}}},\ }\bibfield  {title} {\enquote {\bibinfo {title}
  {{Evolution of entanglement entropy in one-dimensional systems}},}\ }\href
  {\doibase 10.1088/1742-5468/2005/04/P04010} {\bibfield  {journal} {\bibinfo
  {journal} {Journal of Statistical Mechanics: Theory and Experiment}\ }\textbf
  {\bibinfo {volume} {2005}},\ \bibinfo {pages} {04010} (\bibinfo {year}
  {2005})},\ \Eprint {http://arxiv.org/abs/cond-mat/0503393}
  {arXiv:cond-mat/0503393 [cond-mat.stat-mech]} \BibitemShut {NoStop}%
\bibitem [{\citenamefont {{Ginsparg}}(1988)}]{1991appliedcft}%
  \BibitemOpen
  \bibfield  {author} {\bibinfo {author} {\bibfnamefont {Paul}\ \bibnamefont
  {{Ginsparg}}},\ }\bibfield  {title} {\enquote {\bibinfo {title} {{Applied
  Conformal Field Theory}},}\ }\href {\doibase 10.48550/arXiv.hep-th/9108028}
  {\bibfield  {journal} {\bibinfo  {journal} {arXiv e-prints}\ ,\ \bibinfo
  {eid} {hep-th/9108028}} (\bibinfo {year} {1988})},\ \Eprint
  {http://arxiv.org/abs/hep-th/9108028} {arXiv:hep-th/9108028 [hep-th]}
  \BibitemShut {NoStop}%
\bibitem [{\citenamefont {{Oshikawa}}\ and\ \citenamefont
  {{Affleck}}(1997)}]{1997affleck}%
  \BibitemOpen
  \bibfield  {author} {\bibinfo {author} {\bibfnamefont {Masaki}\ \bibnamefont
  {{Oshikawa}}}\ and\ \bibinfo {author} {\bibfnamefont {Ian}\ \bibnamefont
  {{Affleck}}},\ }\bibfield  {title} {\enquote {\bibinfo {title} {{Boundary
  conformal field theory approach to the critical two-dimensional Ising model
  with a defect line}},}\ }\href {\doibase 10.1016/S0550-3213(97)00219-8}
  {\bibfield  {journal} {\bibinfo  {journal} {Nuclear Physics B}\ }\textbf
  {\bibinfo {volume} {495}},\ \bibinfo {pages} {533--582} (\bibinfo {year}
  {1997})},\ \Eprint {http://arxiv.org/abs/cond-mat/9612187}
  {arXiv:cond-mat/9612187 [cond-mat.stat-mech]} \BibitemShut {NoStop}%
\bibitem [{\citenamefont {{Oshikawa}}\ and\ \citenamefont
  {{Affleck}}(1996)}]{1996oshikawa}%
  \BibitemOpen
  \bibfield  {author} {\bibinfo {author} {\bibfnamefont {Masaki}\ \bibnamefont
  {{Oshikawa}}}\ and\ \bibinfo {author} {\bibfnamefont {Ian}\ \bibnamefont
  {{Affleck}}},\ }\bibfield  {title} {\enquote {\bibinfo {title} {{Defect Lines
  in the Ising Model and Boundary States on Orbifolds}},}\ }\href {\doibase
  10.1103/PhysRevLett.77.2604} {\bibfield  {journal} {\bibinfo  {journal}
  {\prl}\ }\textbf {\bibinfo {volume} {77}},\ \bibinfo {pages} {2604--2607}
  (\bibinfo {year} {1996})},\ \Eprint {http://arxiv.org/abs/hep-th/9606177}
  {arXiv:hep-th/9606177 [hep-th]} \BibitemShut {NoStop}%
\bibitem [{\citenamefont {Affleck}\ and\ \citenamefont
  {Ludwig}(1991)}]{affleck1991universal}%
  \BibitemOpen
  \bibfield  {author} {\bibinfo {author} {\bibfnamefont {Ian}\ \bibnamefont
  {Affleck}}\ and\ \bibinfo {author} {\bibfnamefont {Andreas~WW}\ \bibnamefont
  {Ludwig}},\ }\bibfield  {title} {\enquote {\bibinfo {title} {Universal
  noninteger ‘‘ground-state degeneracy’’in critical quantum systems},}\
  }\href@noop {} {\bibfield  {journal} {\bibinfo  {journal} {Physical Review
  Letters}\ }\textbf {\bibinfo {volume} {67}},\ \bibinfo {pages} {161}
  (\bibinfo {year} {1991})}\BibitemShut {NoStop}%
\bibitem [{\citenamefont {{Simmons-Duffin}}(2017)}]{2017bootstrap}%
  \BibitemOpen
  \bibfield  {author} {\bibinfo {author} {\bibfnamefont {David}\ \bibnamefont
  {{Simmons-Duffin}}},\ }\bibfield  {title} {\enquote {\bibinfo {title} {{The
  lightcone bootstrap and the spectrum of the 3d Ising CFT}},}\ }\href
  {\doibase 10.1007/JHEP03(2017)086} {\bibfield  {journal} {\bibinfo  {journal}
  {Journal of High Energy Physics}\ }\textbf {\bibinfo {volume} {2017}},\
  \bibinfo {eid} {86} (\bibinfo {year} {2017})},\ \Eprint
  {http://arxiv.org/abs/1612.08471} {arXiv:1612.08471 [hep-th]} \BibitemShut
  {NoStop}%
\bibitem [{\citenamefont {Burkhardt}\ and\ \citenamefont
  {Cardy}(1987)}]{burkhardt1987surface}%
  \BibitemOpen
  \bibfield  {author} {\bibinfo {author} {\bibfnamefont {TW}~\bibnamefont
  {Burkhardt}}\ and\ \bibinfo {author} {\bibfnamefont {JL}~\bibnamefont
  {Cardy}},\ }\bibfield  {title} {\enquote {\bibinfo {title} {Surface critical
  behaviour and local operators with boundary-induced critical profiles},}\
  }\href@noop {} {\bibfield  {journal} {\bibinfo  {journal} {Journal of Physics
  A: Mathematical and General}\ }\textbf {\bibinfo {volume} {20}},\ \bibinfo
  {pages} {L233} (\bibinfo {year} {1987})}\BibitemShut {NoStop}%
\bibitem [{\citenamefont {Bray}\ and\ \citenamefont
  {Moore}(1977)}]{bray1977critical}%
  \BibitemOpen
  \bibfield  {author} {\bibinfo {author} {\bibfnamefont {AJ}~\bibnamefont
  {Bray}}\ and\ \bibinfo {author} {\bibfnamefont {MA}~\bibnamefont {Moore}},\
  }\bibfield  {title} {\enquote {\bibinfo {title} {Critical behaviour of
  semi-infinite systems},}\ }\href@noop {} {\bibfield  {journal} {\bibinfo
  {journal} {Journal of Physics A: Mathematical and General}\ }\textbf
  {\bibinfo {volume} {10}},\ \bibinfo {pages} {1927} (\bibinfo {year}
  {1977})}\BibitemShut {NoStop}%
\bibitem [{\citenamefont {Cardy}(1996)}]{cardy1996scaling}%
  \BibitemOpen
  \bibfield  {author} {\bibinfo {author} {\bibfnamefont {John}\ \bibnamefont
  {Cardy}},\ }\href@noop {} {\emph {\bibinfo {title} {Scaling and
  renormalization in statistical physics}}},\ Vol.~\bibinfo {volume} {5}\
  (\bibinfo  {publisher} {Cambridge university press},\ \bibinfo {year}
  {1996})\BibitemShut {NoStop}%
\bibitem [{\citenamefont {Sachdev}(1999)}]{sachdev1999quantum}%
  \BibitemOpen
  \bibfield  {author} {\bibinfo {author} {\bibfnamefont {Subir}\ \bibnamefont
  {Sachdev}},\ }\bibfield  {title} {\enquote {\bibinfo {title} {Quantum phase
  transitions},}\ }\href@noop {} {\bibfield  {journal} {\bibinfo  {journal}
  {Physics world}\ }\textbf {\bibinfo {volume} {12}},\ \bibinfo {pages} {33}
  (\bibinfo {year} {1999})}\BibitemShut {NoStop}%
\bibitem [{\citenamefont {{Senthil}}\ \emph {et~al.}(2019)\citenamefont
  {{Senthil}}, \citenamefont {{Son}}, \citenamefont {{Wang}},\ and\
  \citenamefont {{Xu}}}]{2019Chong}%
  \BibitemOpen
  \bibfield  {author} {\bibinfo {author} {\bibfnamefont {T.}~\bibnamefont
  {{Senthil}}}, \bibinfo {author} {\bibfnamefont {Dam~Thanh}\ \bibnamefont
  {{Son}}}, \bibinfo {author} {\bibfnamefont {Chong}\ \bibnamefont {{Wang}}}, \
  and\ \bibinfo {author} {\bibfnamefont {Cenke}\ \bibnamefont {{Xu}}},\
  }\bibfield  {title} {\enquote {\bibinfo {title} {{Duality between (2 + 1) d
  quantum critical points}},}\ }\href {\doibase 10.1016/j.physrep.2019.09.001}
  {\bibfield  {journal} {\bibinfo  {journal} {physrep}\ }\textbf {\bibinfo
  {volume} {827}},\ \bibinfo {pages} {1--48} (\bibinfo {year} {2019})},\
  \Eprint {http://arxiv.org/abs/1810.05174} {arXiv:1810.05174
  [cond-mat.str-el]} \BibitemShut {NoStop}%
\bibitem [{\citenamefont {{Turkeshi}}\ \emph {et~al.}(2021)\citenamefont
  {{Turkeshi}}, \citenamefont {{Biella}}, \citenamefont {{Fazio}},
  \citenamefont {{Dalmonte}},\ and\ \citenamefont
  {{Schir{\'o}}}}]{2021Turkeshi}%
  \BibitemOpen
  \bibfield  {author} {\bibinfo {author} {\bibfnamefont {Xhek}\ \bibnamefont
  {{Turkeshi}}}, \bibinfo {author} {\bibfnamefont {Alberto}\ \bibnamefont
  {{Biella}}}, \bibinfo {author} {\bibfnamefont {Rosario}\ \bibnamefont
  {{Fazio}}}, \bibinfo {author} {\bibfnamefont {Marcello}\ \bibnamefont
  {{Dalmonte}}}, \ and\ \bibinfo {author} {\bibfnamefont {Marco}\ \bibnamefont
  {{Schir{\'o}}}},\ }\bibfield  {title} {\enquote {\bibinfo {title}
  {{Measurement-induced entanglement transitions in the quantum Ising chain:
  From infinite to zero clicks}},}\ }\href {\doibase
  10.1103/PhysRevB.103.224210} {\bibfield  {journal} {\bibinfo  {journal}
  {\prb}\ }\textbf {\bibinfo {volume} {103}},\ \bibinfo {eid} {224210}
  (\bibinfo {year} {2021})},\ \Eprint {http://arxiv.org/abs/2103.09138}
  {arXiv:2103.09138 [quant-ph]} \BibitemShut {NoStop}%
\bibitem [{\citenamefont {{Chen}}\ \emph {et~al.}(2013)\citenamefont {{Chen}},
  \citenamefont {{Gu}}, \citenamefont {{Liu}},\ and\ \citenamefont
  {{Wen}}}]{2013SPTchen}%
  \BibitemOpen
  \bibfield  {author} {\bibinfo {author} {\bibfnamefont {Xie}\ \bibnamefont
  {{Chen}}}, \bibinfo {author} {\bibfnamefont {Zheng-Cheng}\ \bibnamefont
  {{Gu}}}, \bibinfo {author} {\bibfnamefont {Zheng-Xin}\ \bibnamefont {{Liu}}},
  \ and\ \bibinfo {author} {\bibfnamefont {Xiao-Gang}\ \bibnamefont {{Wen}}},\
  }\bibfield  {title} {\enquote {\bibinfo {title} {{Symmetry protected
  topological orders and the group cohomology of their symmetry group}},}\
  }\href {\doibase 10.1103/PhysRevB.87.155114} {\bibfield  {journal} {\bibinfo
  {journal} {\prb}\ }\textbf {\bibinfo {volume} {87}},\ \bibinfo {eid} {155114}
  (\bibinfo {year} {2013})},\ \Eprint {http://arxiv.org/abs/1106.4772}
  {arXiv:1106.4772 [cond-mat.str-el]} \BibitemShut {NoStop}%
\bibitem [{\citenamefont {{Tonielli}}\ \emph {et~al.}(2020)\citenamefont
  {{Tonielli}}, \citenamefont {{Budich}}, \citenamefont {{Altland}},\ and\
  \citenamefont {{Diehl}}}]{2020drivenchern}%
  \BibitemOpen
  \bibfield  {author} {\bibinfo {author} {\bibfnamefont {F.}~\bibnamefont
  {{Tonielli}}}, \bibinfo {author} {\bibfnamefont {J.~C.}\ \bibnamefont
  {{Budich}}}, \bibinfo {author} {\bibfnamefont {A.}~\bibnamefont {{Altland}}},
  \ and\ \bibinfo {author} {\bibfnamefont {S.}~\bibnamefont {{Diehl}}},\
  }\bibfield  {title} {\enquote {\bibinfo {title} {{Topological Field Theory
  Far from Equilibrium}},}\ }\href {\doibase 10.1103/PhysRevLett.124.240404}
  {\bibfield  {journal} {\bibinfo  {journal} {\prl}\ }\textbf {\bibinfo
  {volume} {124}},\ \bibinfo {eid} {240404} (\bibinfo {year} {2020})},\ \Eprint
  {http://arxiv.org/abs/1911.07834} {arXiv:1911.07834 [cond-mat.stat-mech]}
  \BibitemShut {NoStop}%
\bibitem [{\citenamefont {{Ma}}\ and\ \citenamefont {{Wang}}(2022)}]{2022aspt}%
  \BibitemOpen
  \bibfield  {author} {\bibinfo {author} {\bibfnamefont {Ruochen}\ \bibnamefont
  {{Ma}}}\ and\ \bibinfo {author} {\bibfnamefont {Chong}\ \bibnamefont
  {{Wang}}},\ }\bibfield  {title} {\enquote {\bibinfo {title} {{Average
  Symmetry-Protected Topological Phases}},}\ }\href {\doibase
  10.48550/arXiv.2209.02723} {\bibfield  {journal} {\bibinfo  {journal} {arXiv
  e-prints}\ ,\ \bibinfo {eid} {arXiv:2209.02723}} (\bibinfo {year} {2022})},\
  \Eprint {http://arxiv.org/abs/2209.02723} {arXiv:2209.02723
  [cond-mat.str-el]} \BibitemShut {NoStop}%
\end{thebibliography}%

\clearpage
\onecolumngrid
\appendix

\section{Weak Measurements}
\label{app:qsdderivation}

This Appendix presents a derivation of the state evolution under weak measurements. For simplicity, I focus on a two-level system as a toy model. A generalization to the many body cases, Eq.~(\ref{eq:qsdfinite}) and Eq.~(\ref{eq:qsdinfinite}), is straightforward. The derivation follows a methodology similar to that outlined in Ref.~\cite{2002trajectory}.

Consider a two dimensional Hilbert space, with basis states $|0\rangle$ and $|1\rangle$. The two basis states are eigenstates of a Hermitian operator $O$, which has eigenvalues of 0 and 1. A generalized measurement is a partition of unity by non-negative Hermitian operators:
\begin{equation}
    \sum_a A_a^\dagger A_a = \mathbf{1}.
\end{equation}
The probability of obtaining the outcome $a$ in a generic state $| \Psi \rangle$ is given by $\langle \Psi | A_a^\dagger A_a | \Psi \rangle$.

In the two level system, projective measurements in the eigenbasis of $O$ are implemented by
\begin{equation}
    P_0 = \mathbf{1} - O,\quad P_1 =  O,
\end{equation}
which project the state onto either $|0\rangle$ or $|1\rangle$ depending on the measurement outcome. If we want a measurement that only alters the state $| \Psi \rangle$ slightly, i.e. a continuous measurement, we can choose
\begin{equation}
    A_\pm = \sqrt{\frac{1\pm \Gamma}{2}}P_0 + \sqrt{\frac{1\mp \Gamma}{2}}P_1,
\end{equation}
with $\Gamma \in [0,1]$ being the measurement strength. The expression is simply an interpolation between no measurement ($\Gamma=0$) and projective measurement ($\Gamma=1$). When $\Gamma \ll 1$, the post-measurement state $| \Psi' \rangle$ is only weakly altered:
\begin{equation}
| \Psi' \rangle =
\begin{cases}
A_+| \Psi \rangle /\sqrt{p_+}, & \text{if $+$ measured}, \\
A_-| \Psi \rangle /\sqrt{p_-}, & \text{if $-$ measured} ,
\end{cases}
\end{equation}
with the probabilities $p_\pm = \frac{1}{2}[1 \pm \Gamma (1-2\langle O \rangle)]$. Here $\langle O \rangle = \langle \Psi | O | \Psi \rangle$ is the expectation value of $O$ in the pre-measurement state. The change in the state density matrix, up to first order in $\Gamma$ is given by
\begin{equation}
    \delta \rho \approx -\frac{\Gamma}{2}[M,[M,\rho]] + \sqrt{\Gamma} W \{ M,\rho \},
\end{equation}
where the measurement operator $M$ is defined as $M = : O-\langle O \rangle$. $W = \pm 1$ for outcome $+$ and $-$, respectively. This equation immediately implies Eq.~(\ref{eq:qsdfinite}) and Eq.~(\ref{eq:qsdinfinite}).

In the scenario where measurements are performed over an $O(L)$ time, it is important to handle the averaging over the trajectory ensemble with care, particularly when $n>1$ replicas are involved. This is because the measurement action $S_M$ explicitly depends on the expectation value of $O$ within a specific outcome trajectory, i.e., $\langle O \rangle(t) = \mathrm{tr}[O\rho(t)]$, which arises from the definition of the measurement operator $M_t = O-\langle O \rangle(t)$. Therefore, to solve the time evolution of $\rho_n = \overline{\rho^{\otimes n}}$ for $n>1$ (in this study, we focused mostly on $\rho_2$), it is necessary to determine the evolution of $\rho_{n+1}$ simultaneously \cite{2021buchhold}. To handle this, we employ a mean-field approximation to account for the measurement feedback on the time evolution. Specifically, the trajectory-averaged product is approximated as
\begin{equation}
    \overline{\mathrm{tr}(\rho O)\cdot \rho \otimes \rho} \approx \mathrm{tr}(\rho_2 O)\cdot \rho_2,
\end{equation}
which becomes more accurate as the measurement strength $\Gamma$ decreases. This approximation is valid for the purpose of our study, which is to investigate the stability of the critical system against infinitesimal measurement. Accordingly, the trajectory-averaged expectation value $\mathrm{tr}(\rho_2 O)$ is then replaced by the expectation value of $O$ in the unperturbed ground state, which is absorbed by normal ordering of the CFT operators. As a result, all one-point functions in the CFT are set to 0.

\end{document}